\documentclass[12pt,preprint]{aastex}
\usepackage{datetime}
\usepackage{natbib}
\shorttitle{M31/M33 Spectroscopy}

\shortauthors{Massey et al.}

\begin{document}
\setlength{\unitlength}{2em}

\title{A Spectroscopic Survey of Massive Stars in M31 and M33\altaffilmark{*}}

\author{Philip Massey\altaffilmark{1,2},   Kathryn F. Neugent\altaffilmark{1,2}, \and Brianna M. Smart\altaffilmark{1,3}}

\altaffiltext{*}{The spectroscopic observations reported here were obtained at the MMT Observatory, a joint facility of the University of Arizona and the Smithsonian Institution. MMT telescope time was granted by NOAO, through the Telescope System Instrumentation Program (TSIP). TSIP is funded by the National Science Foundation. This paper uses data products produced by the OIR Telescope Data Center, supported by the Smithsonian Astrophysical Observatory.}
\altaffiltext{1}{Lowell Observatory, 1400 W Mars Hill Road, Flagstaff, AZ 86001; phil.massey@lowell.edu; kneugent@lowell.edu.}

\altaffiltext{2}{Dept.\ Physics \& Astronomy, Northern Arizona University, P. O. Box 6010, Flagstaff, AZ 86011-6010. }

\altaffiltext{3}{Research Experience for Undergraduate participant during the summer of 2011.  Current address: Department of Astronomy,  University of Wisconsin-Madison, 2535 Sterling Hall, 475 N. Charter Street, Madison, WI 53706-1582; bsmart@astro.wisc.edu. }


\begin{abstract}

We describe our spectroscopic follow-up to the Local Group Galaxy Survey (LGGS) photometry of 
M31 and M33.  We have
obtained new spectroscopy of 1895 stars, allowing us to classify 1496 of them for the first time.  Our study has identified many foreground stars, and established membership for hundreds of early- and mid-type supergiants.  We have also found 9 new candidate Luminous Blue Variables and a previously unrecognized Wolf-Rayet star.  We republish the LGGS M31 and M33 catalogs with improved coordinates and including spectroscopy from the literature and our new results. The spectroscopy in this paper is responsible for the vast majority of the stellar classifications in these two nearby spiral neighbors.  The most luminous (and hence massive) of the stars in our sample are early-type B supergiants, as expected; the more massive O stars will be fainter visually, and thus mostly remain unobserved so far.  The majority of the unevolved stars in our sample are in the 20-40$M_\odot$ range.

\end{abstract}


\section{Introduction}

The galaxies of the Local Group serve as our astrophysical laboratories, where we can test massive star evolutionary models as a function of metallicity.
On the main-sequence, massive stars are so hot and luminous that radiation pressure, acting on ionized metal atoms, drives off their outer layers.  The mass-loss rates are high enough that they affect the evolution of the star\footnote{The inclusion of non-homogeneities  (``clumping") in the winds of hot massive stars has lowered the inferred mass-loss rates by $\sim 3\times$ from what we believed a decade ago \cite[e.g.,][]{OwockiCohen06,PulsClumping}, but the rates are still high enough to significantly affect the evolution of these stars; see, for example, \citet{Sylvia}.}.  Since they depend upon the initial metallicity of the gas out of which the stars formed, we see dramatic differences in the relative numbers of different types of evolved massive stars throughout the Local Group.  For instance, we observe an order of magnitude change in the relative number of red supergiants (RSGs) to Wolf-Rayet stars (WRs) from the SMC to M31 \citep{MasseyARAA,MasseyPots}, as first predicted by \citet{Maeder80}.    We observe a similarly dramatic change in the relative number of WN- and WC-type WRs \citep{NeugentM31} with metallicity, as well as changes in the relative proportions of early- and  late-type WNs, and early- and late-type WCs \citep[see, e.g.,][]{MasseyPots}.  Identifying these populations allows us to compare observations with the predictions of stellar evolutionary models, discovering where the agreement is good, and where it is not \cite[see, e.g.,][]{DroutM31,DroutM33,NeugentSMC,NeugentM33,NeugentLMC,NeugentM31}. However, until now very little has been done on the much harder problem of characterizing the relatively {\it unevolved} massive star populations of these galaxies.

The Local Group Galaxy Survey (LGGS) provides {\it UBVRI} plus interference-image photometry of luminous stars in the spiral galaxies M31 and M33, along with those found in seven dwarf systems currently forming massive stars (IC 10, NGC~6822, WLM,  Sextans A and B, Pegasus, and Phoenix) with the intent of serving as a starting point for systematic exploration of the stellar content of our nearest neighbors beyond the
Magellanic Clouds \citep{LGGSI,LGGSII,LGGSIII,LGGSIV}.  However, such photometric studies require spectroscopic follow-up to characterize these systems' massive star populations for two reasons.   First, on the main-sequence, the massive stars are so hot that most of their energy is in the far-ultraviolet, and their optical fluxes fall well onto the tail of the Rayleigh-Jeans distribution.  Thus, there is little difference in the optical colors of such stars with effective temperature.  Yet, the bolometric corrections needed to get a star's total luminosity ($L$), and hence an inferred mass ($m$), are a very sensitive function of effective temperature, a point strongly emphasized and demonstrated by \citet{MasseyPyke}.  \citet{MasseyGilmore} finds that even using the reddening-free Johnson Q-index (Johnson \& Morgan 1953), a change of 0.1~mag in the color corresponds to a difference from 30,000 K to 50,000 K, with a resounding 0.6~dex change in $\log L/L_\odot$, or $\sim 0.3$ dex in $\log m$; i.e., a factor of 2 in the mass.

\begin{figure}
\epsscale{0.5}
\plotone{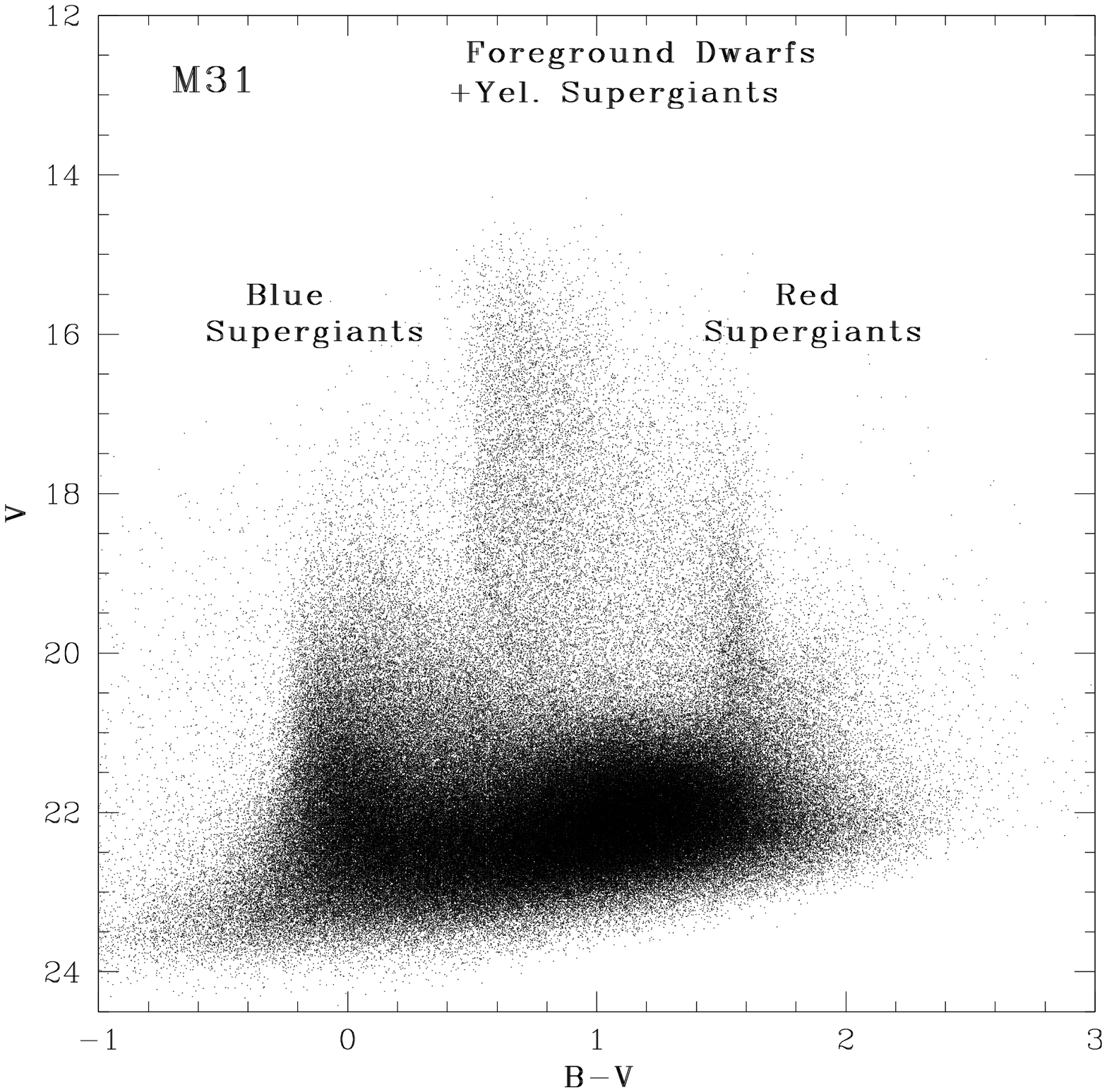}
\plotone{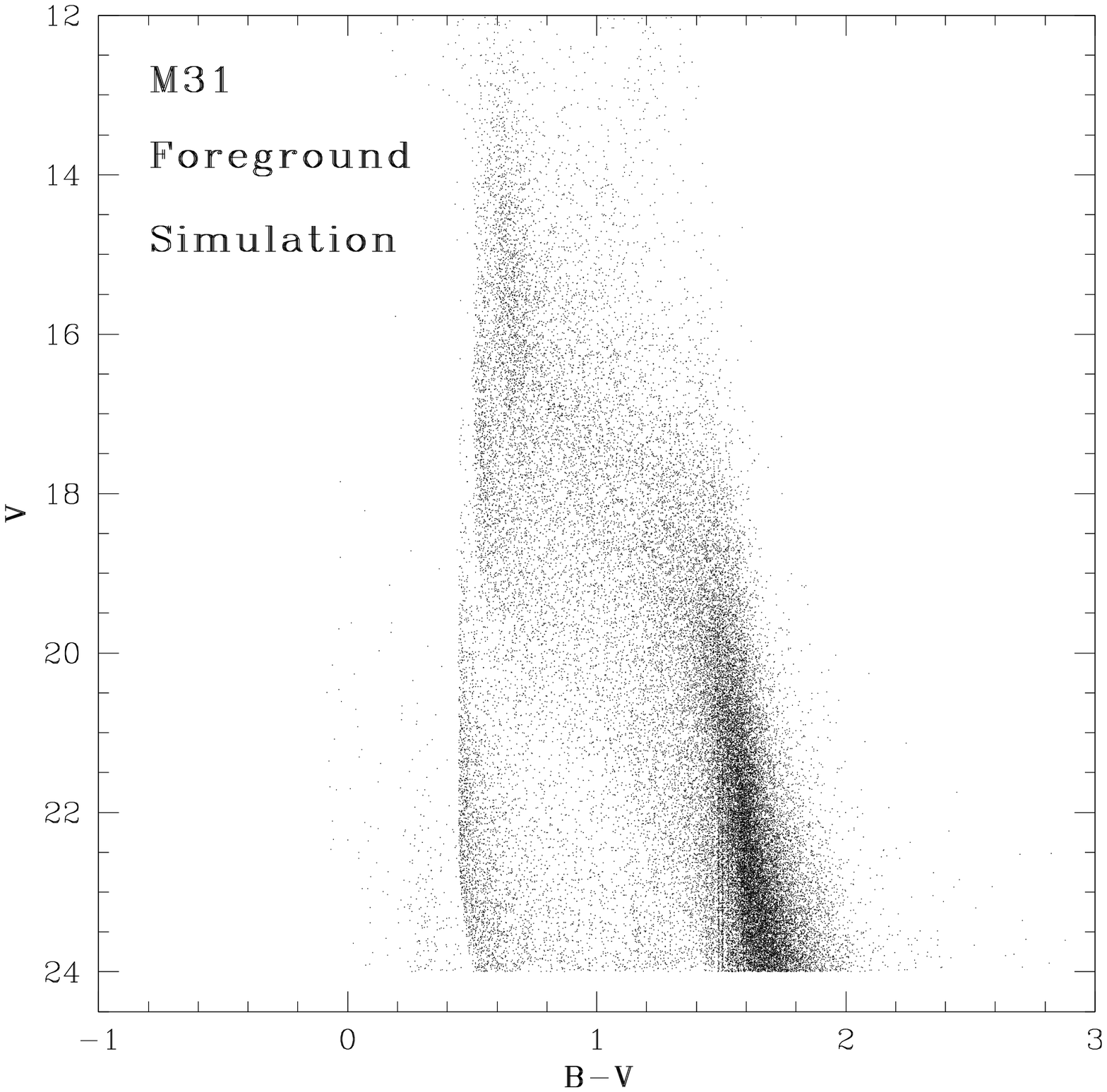}
\caption{\label{fig:m31cmd} Color-magnitude diagram M31.  {\it Top.} The figure shows the observed color-magnitude diagram for M31 from the LGGS.  {\it Bottom.} The figure shows the expected foreground contamination at the same Galactic longitude ($l=121\fdg2$), latitude ($b=-21\fdg6$), and surface area (2.2 deg$^2$) as for the top figure, as computed from the Besan\c{c}on model \citep{Besancon}.
}

\end{figure}

\begin{figure}
\epsscale{0.5}
\plotone{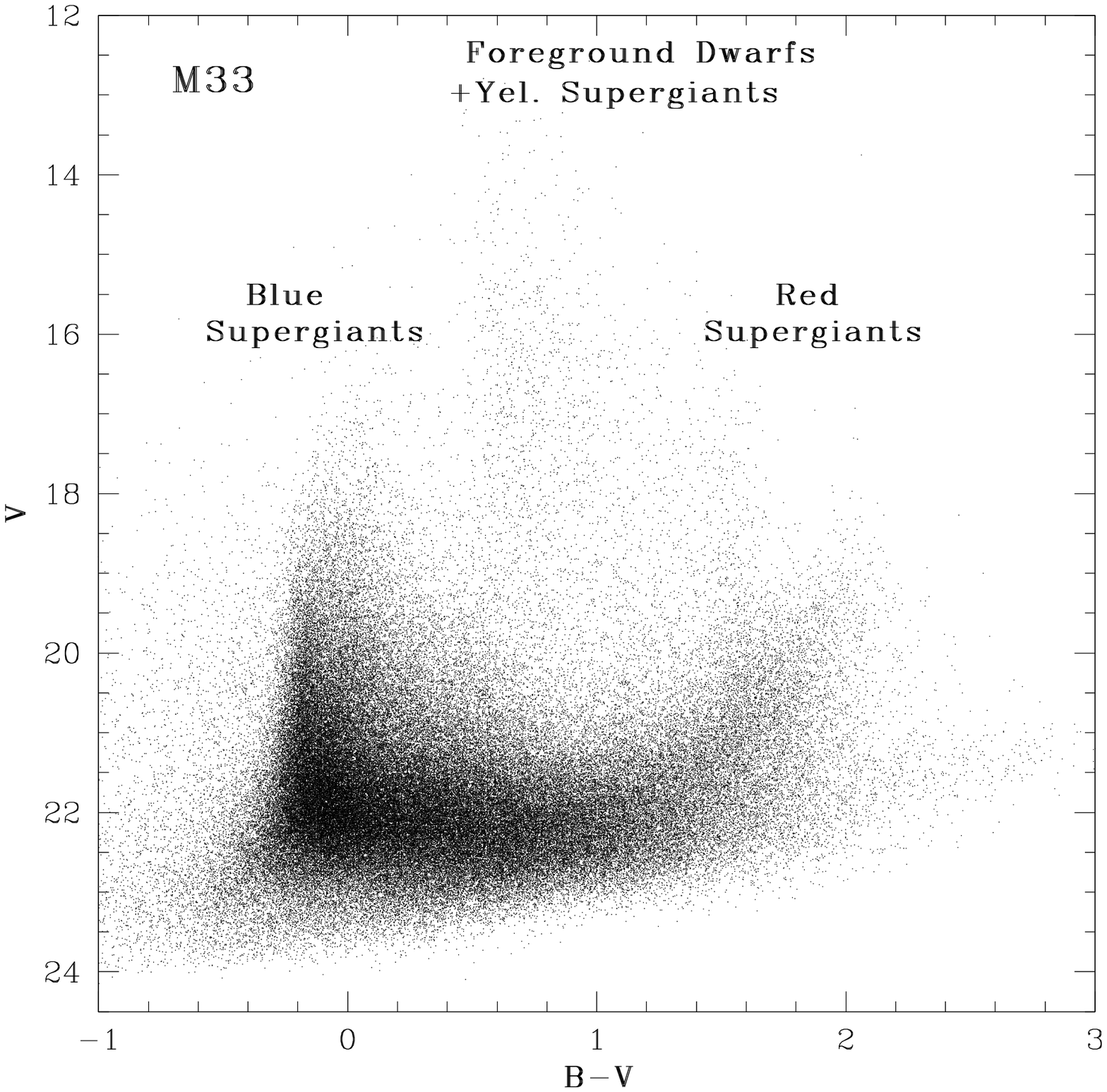}
\plotone{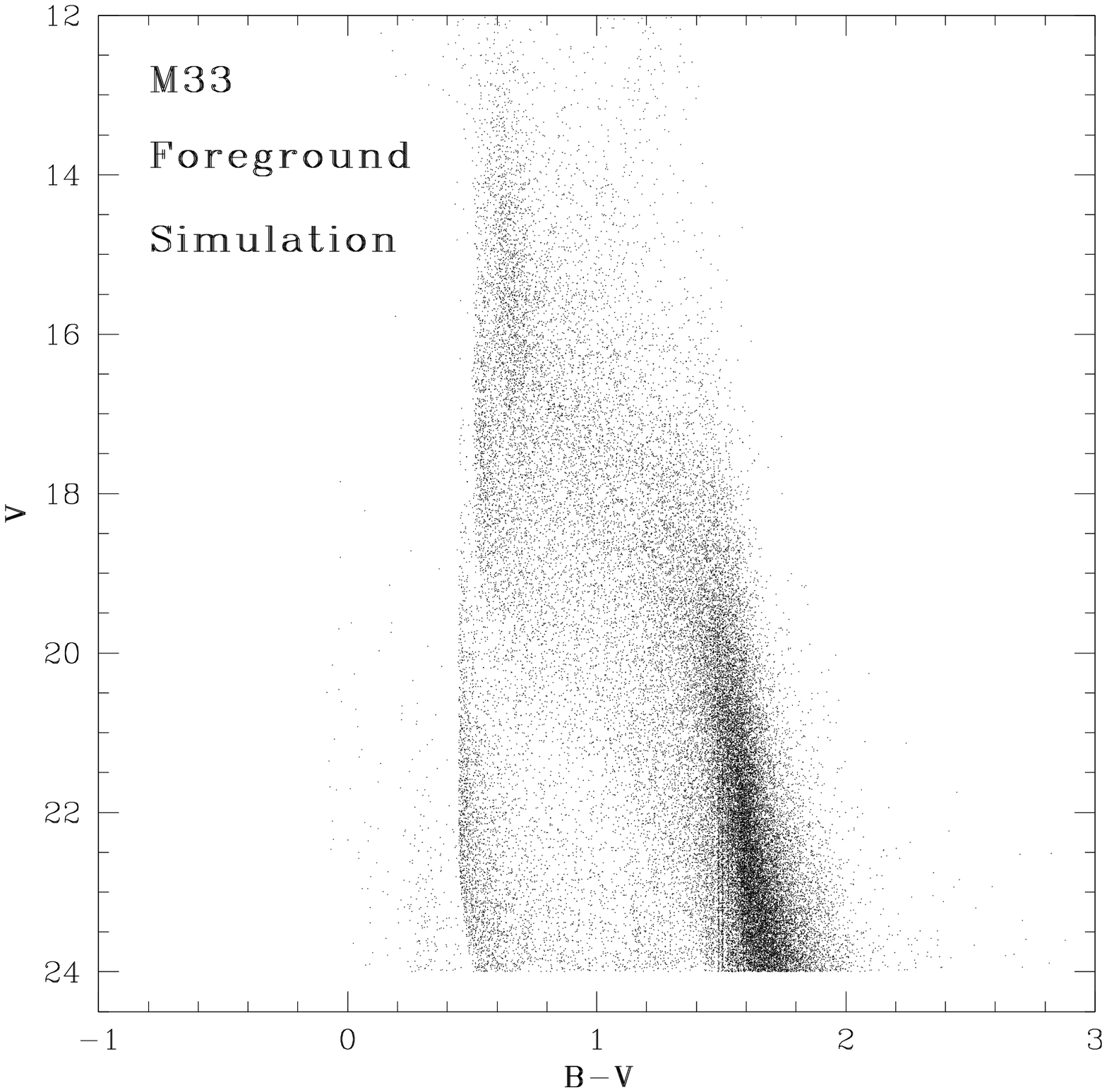}
\caption{\label{fig:m33cmd} Color-magnitude diagram M33.  {\it Top.} The figure shows the observed color-magnitude diagram for M33 from the LGGS.   {\it Bottom.} The figure shows the expected foreground contamination at the same Galactic longitude ($l=133\fdg6$), latitude ($b=-31\fdg3$), and surface area (0.8 deg$^2$) as for the top figure,
as computed from the Besan\c{c}on model \citep{Besancon}.
}

\end{figure}

The second reason has to do with foreground contamination.  We illustrate this in Figures~\ref{fig:m31cmd} and \ref{fig:m33cmd}, where we compare the observed color-magnitude diagrams (CMDs) of the M31 and M33 fields
from the LGGS with that expected by foreground contamination using the Besan\c{c}on Milky Way stellar population synthesis model \citep{Besancon}.  
The foreground contamination is small for the blue supergiants (OB stars), where their very blue colors ($B-V<0$) distinguish them from common foreground
stars in the right magnitude range.   But, as a massive star evolves, it may pass through a yellow supergiant (YSG) phase and then on to a red supergiant (RSG) phase.
 The lifetimes of stars in the YSG phase are {\it very} short (tens of thousands of years or even less), and studies of the numbers and luminosities of such stars have proven to be very useful tests of stellar evolutionary theory \citep[see, e.g.,][]{DroutM31,DroutM33,NeugentSMC,NeugentLMC}, but in this region of the HRD foreground contamination is overwhelming. Using radial velocities,  \citet{DroutM31,DroutM33} found only a few percent of the stars in this region of the CMD were actually yellow supergiants.   The rest were yellow dwarfs in our own galaxy, as evidenced by their radial velocities.  The contamination for the RSGs ($B-V>1.4$) is large for all but the coolest and reddest stars, although foreground dwarfs can be weeded out using two-color diagrams  \citep{MasseyRSG,DroutM33,KateRunaway}. 
 
In this paper we present new spectroscopy for 1895 stars in the LGGS catalog of M31 and M33 stars, of which 1496 are classified for the first time.  To place this in context, Massey et al.\ (2006) provided cross-identifications to previous spectroscopy of stars in these galaxies, a list that include only 172 stars in M31 and 361 in M33. \citep[See Tables 8 and 9 in][]{LGGSI}.    Since that time there have been a variety of studies which have performed spectroscopic follow-ups of the LGGS:  \citet{LGGSIII} classified 42 stars in M31 and 80 in M33 as part of a survey of H$\alpha$ emission-lined stars, \citet{CordinerI,CordinerII,CordinerIII} presented new spectral types for 36 M31 stars and one M33 star as part of a study of diffuse interstellar bands.  Using new interference-filter imaging and follow-up spectroscopy, 
\citet{NeugentM33} and \citet{NeugentM31} identified 107 new WRs in M31 and 55 new WRs in M33, along with revising many older WR types, and discovering many new non-WR interesting stars.  Thus prior to the current study, roughly 850 stars were classified in M31 and M33;  the addition of the $\sim$1500 new spectral types we present here thus results in a three fold increase in the number of M31 and M33 stars with spectral types\footnote{\citet{DroutM31} and \citet{DroutM33} relied upon radial velocities of $\sim 2900$ targets in M31 and 1300 in M33 to identify $\sim$100 YSGs in each galaxy, as well as 189 RSGs in M33; these did not include actual classifications, but represent a substantial gain in our knowledge of the stellar populations of these galaxies.}.   As part of the current paper, we reissue the LGGS catalogs of M31 and M33 including both our new spectral types and those that have been published since they were originally released. 

We expect that these new spectral types will help workers in many fields, enabling (for example) studies of the reddening laws in these galaxies \citep[e.g.,][]{ClaytonM31}, stellar chemical abundance studies \citep[e.g.,][]{Fabio07},  and the determination of fundamental stellar parameters at non-solar abundances \citep[e.g.,][]{MasseyZangari}. 

\section{The Data}

\subsection{The Sample}

Most of the 1895 stars that we classify here were selected from the LGGS based upon their colors and magnitudes.  The primary goal was to isolate the OB population, but
because there is some spread in the average reddening \citep[see][]{MAC86,MasseyPyke,LGGSII} we relied heavily on the Johnson reddening-free index $Q=(U-B)-0.72(B-V)$. Our photometric criteria were designed for obtaining
spectra of OB stars with absolute visual magnitudes brighter than $M_V=-5.5$, roughly corresponding to 
50$M_\odot$ for the youngest O stars, and to 25$M_\odot$ for older (5 Myr) O stars.  Such stars will have $V=19.5$ or
brighter.   The photometric criteria we adopted are given in Table~\ref{tab:sample}.  Generally for the fainter stars ($19.5\geq V > 18.0$) we insisted on a reddening-free index corresponding to an effective temperature of $\sim$23,000~K or hotter \citep[see, e.g., Table 2 in][based upon \citealt{Kurucz79,Kurucz92}]{MasseyGilmore},  i.e., roughly a spectral type of B2.5 or earlier \citep[see Table 3 in][]{Canary}.  However, we imposed the additional restriction that both $U-B$ and $B-V$ colors must be reasonable as well, as shown by the other photometric criteria.  For the brighter stars, we relaxed the requirements on $Q$ as there were fewer stars, and identifying B and A supergiants would be useful.  In order
to avoid the 
plethora of foreground dwarfs that would contaminate the latter sample by mid-A or early-F type, we imposed
 $U-B$ cut-off as a function of $B-V$.  

\begin{figure}
\epsscale{0.48}
\plotone{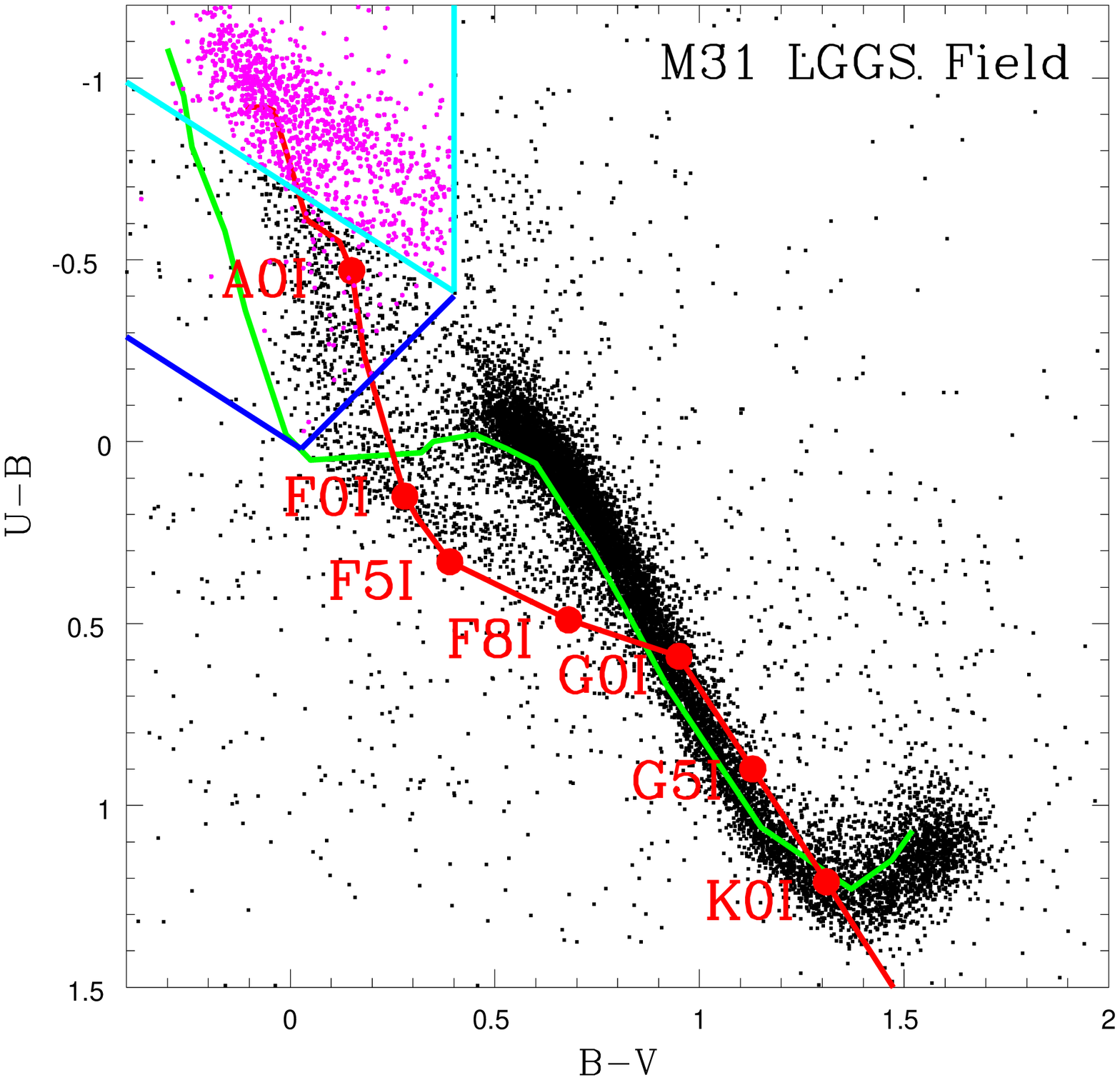}
\plotone{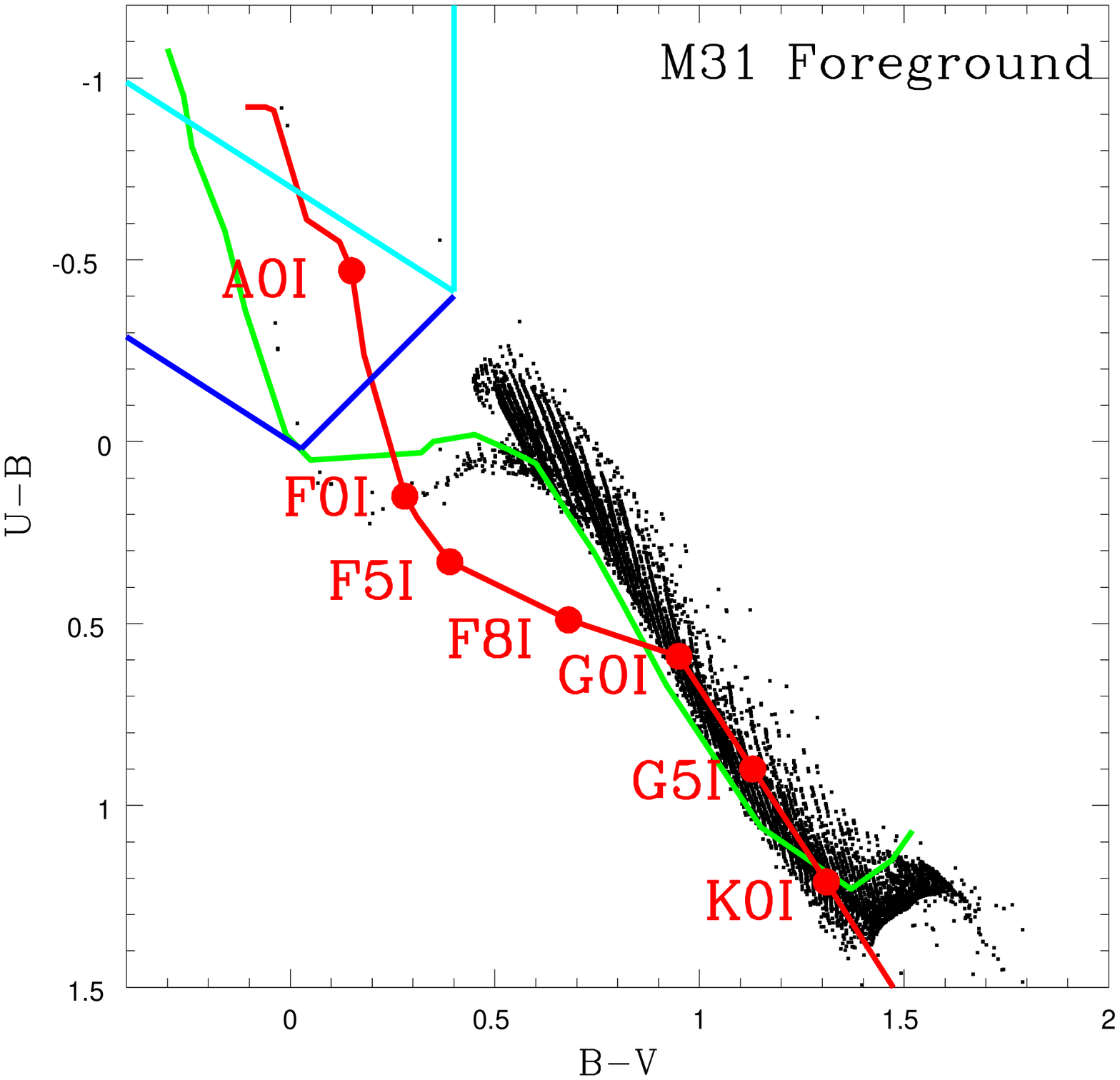}
\caption{\label{fig:m31sel} M31 two-color selection.  {\it Left.} The figure shows the observed two-color magnitude diagram for stars seen towards M31 (black points) from the LGGS for stars with $V\leq19.5$.  The expected two-color relationship for foreground dwarfs is shown in green, while those of supergiants in M31 are shown in red,
where the latter has been adjusted for reddening, and are marked with the expected spectral types.    The magenta filled circles denote the stars that met the photometric criteria given in Table~\ref{tab:sample}.  The
photometric criteria for the fainter stars ($V>18$) are denoted by the two cyan lines, with the additional ``relaxed"
region for brighter stars ($V<18$) shown by the two  blue lines.  {\it Right.}  The figure shows the expected foreground contamination at the same Galactic longitude ($l=121\fdg2$), latitude ($b=-21\fdg6$), and surface area (2.2 deg$^2$) as for the top figure, as computed from the Besan\c{c}on model \citep{Besancon}.
}
\end{figure}

\begin{figure}
\epsscale{0.48}
\plotone{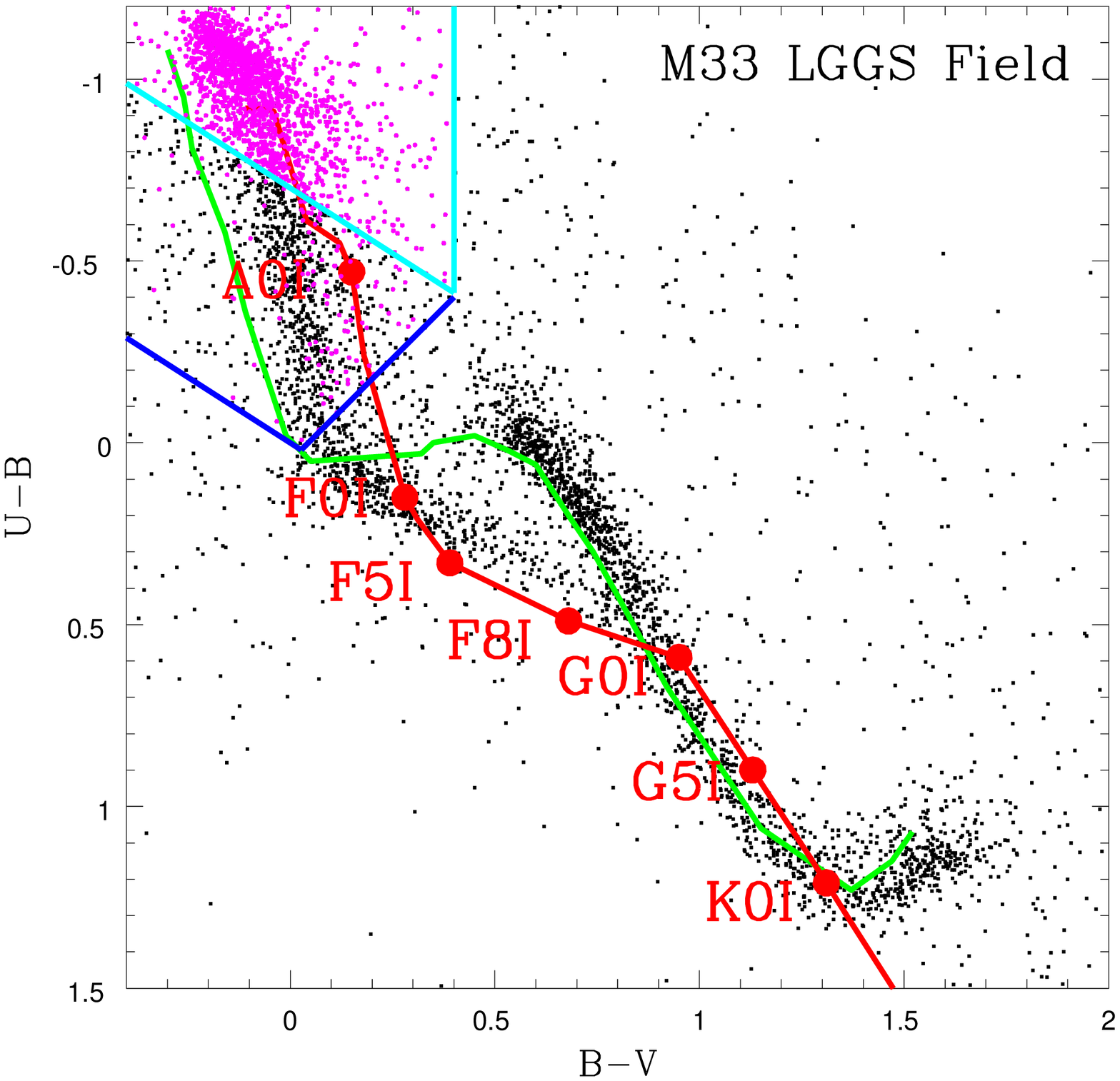}
\plotone{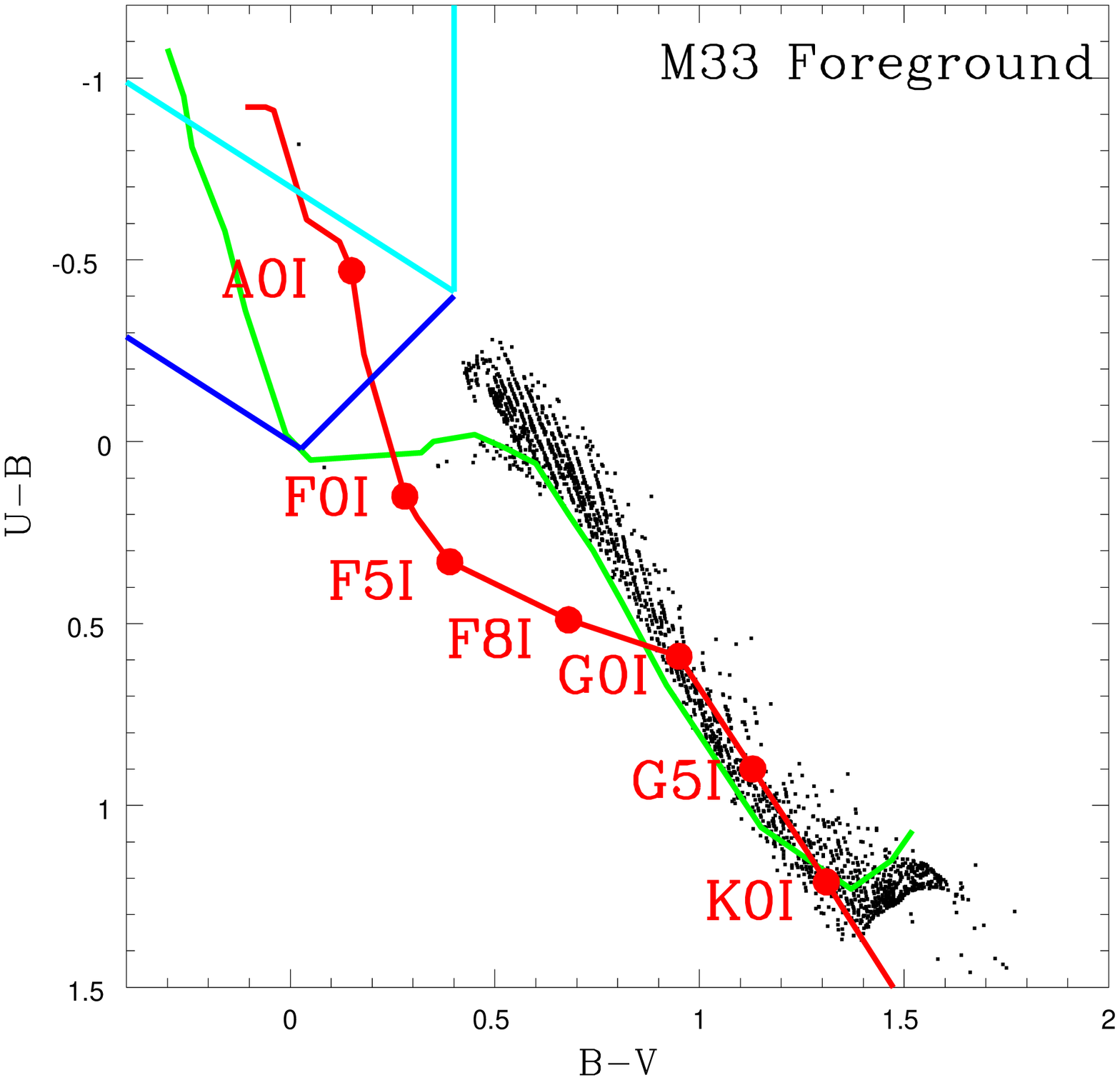}
\caption{\label{fig:m33sel} M33 two-color selection.   {\it Left.} The figure shows the observed two-color magnitude diagram for stars seen towards M33 (black points) from the LGGS for stars with $V\leq19.5$.  The expected two-color relationship for foreground dwarfs is shown in green, while those of supergiants in M33 are shown in red,
where the latter has been adjusted for reddening, and are marked with the expected spectral types.    The magenta filled circles denote the stars that met the photometric criteria given in Table~\ref{tab:sample}.  The
photometric criteria for the fainter stars ($V>18$) are denoted by the two cyan lines, with the additional ``relaxed"
region for brighter stars ($V<18$) shown by the two  blue lines.  {\it Right.} The figure shows the expected foreground contamination at the same Galactic longitude ($l=133\fdg6$), latitude ($b=-31\fdg3$), and surface area (0.8 deg$^2$) as for the top figure,
as computed from the Besan\c{c}on model \citep{Besancon}.
}
\end{figure}

The effect of these selection criteria are shown in
Figures~\ref{fig:m31sel} and \ref{fig:m33sel}.  The black points are the data from the LGGS, with the only restriction that $V\leq 19.5$.  Stars which meet the photometric criteria in Table~\ref{tab:sample} are shown by the magenta filled circles.  The majority of such stars are found in the upper left region, bound by $B-V=0.4$ (vertical cyan line) and $Q<0.7$ (diagonal cyan line).  For the brighter stars ($V\leq 18.0$) we also 
included potential B-A supergiants; these are the stars found in the region bounded by the two blue lines,
representing $Q\leq 0.0$ and $U-B\leq 0.05-1.125(B-V)$.  The latter criteria was imposed to avoid the large number of foreground stars, as demonstrated by the right panels in each figure.  Note that there is very little, if any, expected contamination after applying our photometric criteria.
 
Once the stars were selected photometrically, an additional ``isolation" criteria was imposed.  We considered
the contaminating flux from all other neighboring stars (using the complete LGGS catalog) 
assuming 1\farcs5 seeing.  We required that the contamination be less than 20\% at $B$, although stars
with contamination between 10\% and 20\% were assigned a lower priority in the assignments.  For
M31 this took our original photometrically-selected sample of  1163 stars and reduced it to 517, i.e., only
44\% of the stars were sufficiently isolated.  For M33, this took our photometrically-selected sample of 
1867 stars and reduced it to 619 (i.e., only 33\% were sufficiently isolated)\footnote{We acknowledge that using an isolation criterion may introduce certainly biases into subsequent population studies, but there is little we can do about
that currently.  Also note that we explicitly include a crowding
criteria in our revised LGGS catalogs; the details are not identical to what we have just described.}. 

In addition to the primary program, we also included a number of other targets of potential interest, but at lower priority in assigning fibers.  These include the M31 and M33 confirmed (and suspected) yellow 
supergiants (ranks 1 and 2) from \citet{DroutM31} and \citet{DroutM33}, along with other previously unobserved yellow supergiant candidates selected as described in those papers.  
The latter we expect to be heavily dominated by foreground dwarfs.  This brought the
total potential stars to observe to 3885 for M31, and 1785 for M33.

\subsection{Observations and Reductions}

All of the observations on which our new spectroscopy is based come from the 6.5-m MMT telescope used with the 300 fiber positioner Hectospec \citep{Hecto}\footnote{This is also the telescope/instrument combination used for spectroscopy of the $\sim 2900$ sources
from \citet{DroutM31,DroutM33} as well as the WR candidates observed by \citet{NeugentM33} and \citet{NeugentM31}.}.
This instrument provides a 1$^\circ$ field-of-view, with a choice of two gratings: (1) a 270 line mm$^{-1}$
blazed at 5000~\AA,  which covers the entire visible region (3650-9200~\AA) at once with a resolution of $\sim$5~\AA, and (2) a 600 line mm$^{-1}$ grating blazed 6000~\AA, which covers 2300~\AA\ at one time at a resolution of $\sim$2~\AA.  
The fibers have a diameter of 1\farcs5.  The majority of the new observations reported here
were obtained with the higher dispersion grating nominally centered at 4800~\AA\ but actually covering 3700-6000~\AA.  Some additional new spectral types are those of the non-WRs found by Neugent \& Massey (2011) and Neugent et al.\ (2012a); these data were obtained with the lower-dispersion grating, with details reported in that paper. 

Observations with Hectospec are made in a unique ``self-staffed" queue mode.  The observers are present for specific nights corresponding to the amount of time assigned by the telescope allocation committee, but they obtain observations guided by the
schedule provided by the queue manager.  The schedule is juggled so that at least some of the
objects are the observer's own.  The great advantage of this scheme is that weather losses are distributed amongst all of the
participants.    The first year of observations were obtained during the Fall of 2009, with nominally 3 nights assigned (2009B-0149), but poor weather during the semester meant we obtained only 4 fields in M31 and 1 field in M33.   Each observation consisted of $3\times45$ mins exposures. The second year (Fall 2010), 2 nights were nominally assigned (NOAO 2010B-0260), allowing us to observe 3 additional fields in M33, 
two with 3$\times45$ mins exposures and one with $3\times40$ mins exposures. 

The spectra were all run through the standard  SAO/CfA Hectospec IRAF\footnote{IRAF is distributed by NOAO, which is operated by AURA under cooperative agreement with the NSF.} pipeline \citep{Mink} by Susan Tokarz of the OIR Telescope Data Center; an overview of the process was given by \citet{DroutM31}.   Note that SAO Telescope Data Center is currently in the process of reprocessing all Hectospec
data, and plan a public release in late 2015; we have given permission for the data contained in the present
paper to be included in this archive, and so as to avoid multiple versions floating around the astronomical community, we are not making our spectra separately available here.

\subsection{Spectral Classifications}

The new spectral types were classified on the standard MK system with reference to \cite{WF90}, \citet{Jaschek2}, \citet{MorganAbt}, and \citet{JacobyHunter}.  The initial pass was done by B. M. S., with a careful reality 
check by K. F. N.
Although the basic tenets of stellar spectral classification are well described elsehwere, we briefly summarize our procedure here in case it proves useful to others.  

For the initial classification into spectral class, we followed this algorithm.  If TiO molecular bands were present, the star was of M-type.  (There were not many of these among our stars of course, given our photometric selection criteria; they were mainly introduced into the observing sample by the attempt to be inclusive for yellow supergiants.) Otherwise, the Balmer lines H$\delta$, H$\gamma$, and H$\beta$ were used as preliminary indicators of spectral type. If they were dominant, the star is early-type (O, B, or A). If that was true, next we checked for the presence of He\,{\sc i} and/or He\,{\sc ii}.  If neither were present, the star was of A-type.  If He\,{\sc i} was present, but not He~{\sc ii}, then the star was of B-type.  If He\,{\sc ii} was present, then the star was of O-type.  If instead the Balmer lines were relatively weak, the G band at $\lambda 4308$ was used as the primary discriminant.   If the G-band was completely absent (and the Balmer lines weak), the star was dismissed as an F-type foreground dwarf.  If the G-band is present, but weaker than H$\gamma$, then the star was considered of F-type.  If it was comparable to (or stronger) than H$\gamma$, then the star was of G-K.  If so, we next compared H$\gamma$ to Fe\,{\sc i}~$\lambda 4325$.   If H$\gamma$ was equal or stronger to Fe\,{\sc i}, then it was a G star; if not, then it was of K type.   We then classified the spectrum and assigned a luminosity type (and hence membership) as described below.   To briefly summarize:

\begin{picture}(19.000000,17.000000)(-5.000000,-17.000000)
\put(0.0000,-2.0000){\line(1,1){2.0000}}
\put(0.0000,-2.0000){\line(1,-1){2.0000}}
\put(4.0000,-2.0000){\line(-1,-1){2.0000}}
\put(4.0000,-2.0000){\line(-1,1){2.0000}}
\put(0.0000,-4.0000){\makebox(4.0000,4.0000)[c]{\shortstack[c]{
TiO lines?
}}}
\put(4.0000,-1.4000){\makebox(0,0)[lt]{Yes}}
\put(2.6000,-4.0000){\makebox(0,0)[lb]{No}}
\put(4.0000,-2.0000){\vector(1,0){1.0000}}
\put(7.0000,-2.0000){\oval(4.0000,2.0000)}
\put(5.0000,-3.0000){\makebox(4.0000,2.0000)[c]{\shortstack[c]{
M star
}}}
\put(2.0000,-4.0000){\vector(0,-1){1.0000}}
\put(0.0000,-7.0000){\line(1,1){2.0000}}
\put(0.0000,-7.0000){\line(1,-1){2.0000}}
\put(4.0000,-7.0000){\line(-1,-1){2.0000}}
\put(4.0000,-7.0000){\line(-1,1){2.0000}}
\put(0.0000,-9.0000){\makebox(4.0000,4.0000)[c]{\shortstack[c]{
Balmer lines \\
dominate?
}}}
\put(4.0000,-6.4000){\makebox(0,0)[lt]{No}}
\put(2.6000,-9.0000){\makebox(0,0)[lb]{Yes}}
\put(4.0000,-7.0000){\vector(1,0){1.0000}}
\put(5.0000,-7.0000){\line(1,1){2.0000}}
\put(5.0000,-7.0000){\line(1,-1){2.0000}}
\put(9.0000,-7.0000){\line(-1,-1){2.0000}}
\put(9.0000,-7.0000){\line(-1,1){2.0000}}
\put(5.0000,-9.0000){\makebox(4.0000,4.0000)[c]{\shortstack[c]{
G-band $<$\\
H$\gamma$?
}}}
\put(9.0000,-6.4000){\makebox(0,0)[lt]{Yes}}
\put(7.6000,-9.0000){\makebox(0,0)[lb]{No}}
\put(9.0000,-7.0000){\vector(1,0){1.0000}}
\put(12.0000,-7.0000){\oval(4.0000,2.0000)}
\put(10.0000,-8.0000){\makebox(4.0000,2.0000)[c]{\shortstack[c]{
F star
}}}
\put(7.0000,-9.0000){\vector(0,-1){1.0000}}
\put(5.0000,-12.0000){\line(1,1){2.0000}}
\put(5.0000,-12.0000){\line(1,-1){2.0000}}
\put(9.0000,-12.0000){\line(-1,-1){2.0000}}
\put(9.0000,-12.0000){\line(-1,1){2.0000}}
\put(5.0000,-14.0000){\makebox(4.0000,4.0000)[c]{\shortstack[c]{
H$\gamma <$\\
Fe\,{\sc i} $\lambda 4325$
}}}
\put(9.0000,-11.4000){\makebox(0,0)[lt]{Yes}}
\put(7.6000,-14.0000){\makebox(0,0)[lb]{No}}
\put(9.0000,-12.0000){\vector(1,0){1.0000}}
\put(12.0000,-12.0000){\oval(4.0000,2.0000)}
\put(10.0000,-13.0000){\makebox(4.0000,2.0000)[c]{\shortstack[c]{
K star
}}}
\put(7.0000,-14.0000){\vector(0,-1){1.0000}}
\put(7.0000,-16.0000){\oval(4.0000,2.0000)}
\put(5.0000,-17.0000){\makebox(4.0000,2.0000)[c]{\shortstack[c]{
G star
}}}
\put(2.0000,-9.0000){\vector(0,-1){1.0000}}
\put(0.0000,-12.0000){\line(1,1){2.0000}}
\put(0.0000,-12.0000){\line(1,-1){2.0000}}
\put(4.0000,-12.0000){\line(-1,-1){2.0000}}
\put(4.0000,-12.0000){\line(-1,1){2.0000}}
\put(0.0000,-14.0000){\makebox(4.0000,4.0000)[c]{\shortstack[c]{
He present?
}}}
\put(0.0000,-11.4000){\makebox(0,0)[rt]{Yes}}
\put(2.6000,-14.0000){\makebox(0,0)[lb]{No}}
\put(0.0000,-12.0000){\vector(-1,0){1.0000}}
\put(-5.0000,-12.0000){\line(1,1){2.0000}}
\put(-5.0000,-12.0000){\line(1,-1){2.0000}}
\put(-1.0000,-12.0000){\line(-1,-1){2.0000}}
\put(-1.0000,-12.0000){\line(-1,1){2.0000}}
\put(-5.0000,-14.0000){\makebox(4.0000,4.0000)[c]{\shortstack[c]{
He II present?
}}}
\put(-2.4000,-10.0000){\makebox(0,0)[lt]{Yes}}
\put(-2.4000,-14.0000){\makebox(0,0)[lb]{No}}
\put(-3.0000,-10.0000){\vector(0,1){1.0000}}
\put(-3.0000,-8.0000){\oval(4.0000,2.0000)}
\put(-5.0000,-9.0000){\makebox(4.0000,2.0000)[c]{\shortstack[c]{
O star
}}}
\put(-3.0000,-14.0000){\vector(0,-1){1.0000}}
\put(-3.0000,-16.0000){\oval(4.0000,2.0000)}
\put(-5.0000,-17.0000){\makebox(4.0000,2.0000)[c]{\shortstack[c]{
B star
}}}
\put(2.0000,-14.0000){\vector(0,-1){1.0000}}
\put(2.0000,-16.0000){\oval(4.0000,2.0000)}
\put(0.0000,-17.0000){\makebox(4.0000,2.0000)[c]{\shortstack[c]{
A star
}}}
\end{picture}

Of course, in addition to the stars with classical MK spectral types, our spectroscopy identified candidate Luminous Blue Variable stars  and even a new Wolf-Rayet star as described in more detail below. 

\section{Results}

\subsection{New Members}
In Table~\ref{tab:newmembers} we list the number of spectroscopically confirmed members of M31 and M33 as a function of type, denoting the number that are classified here for the first time.   There is some overlap between the categories, i.e., a star might have an uncertain spectral type of A9-F0~I, in which case it will be counted in each category.  Alternatively, a star might be counted twice because its spectral type is composite, i.e., WR+O.   Nevertheless, these numbers are representative of the state of our knowledge of the resolved stellar populations of our closest spiral neighbors.

Many of the stars shown by \citet{DroutM31} and \citet{DroutM33} to be YSG members of M31 and M33 have subsequently been reclassified with MK types in this paper, and thus although no new ``generic" YSGs are listed in Table~\ref{tab:newmembers}, that is somewhat deceptive as these stars now are included under the more exact F~I or G~I headings.  Similarly, many of the RSGs identified by \citet{MasseySilva} were also classified in that paper.  

We discuss each of these groups below:

\paragraph{\bf Candidate Luminous Blue Variables (cLBVs):}  \citet{HS} identified several blue, luminous stars in M31 and M33 that displayed spectacular photometric variability during a 40 year period; these stars, along with the Galactic and Magellanic Cloud prototypes P~Cygni, S~Doradus, and $\eta$~Car are now called Luminous Blue Variables, following the suggestion of \citet{Conti84}.   Their spectra are variable, with a particular LBV sometimes showing an emission-line spectrum (either Ofpe/WN9 or dominated by permitted and forbidden Fe\,{\sc ii} line emission along with He\,{\sc ii} emission, or a P~Cygni like spectra), or in a cooler state due either to a larger radius induced by the star's proximity to the Eddington limit \citep{GrohAGCar}, or the formation of a pseudo-photosphere in the stellar wind, leading to an F supergiant like spectrum  \citep[see, e.g.,][]{MasseySDor}. 
Over the years, a number of other stars have been identified in M31 and M33 that have spectra indistinguishable from those of the classical LBVs.  Some were observed spectroscopically because the stars were known to have emission as shown by narrow-band H$\alpha$ imaging \citep[e.g.,][]{King,LGGSIII}, or due to UV-excess \citep{UIT}, or simply discovered ``accidentally" as part of spectroscopic surveys, such as that of \citet{M31PCyg} and the present paper.  Many of these ``candidate" LBVs have now been  shown to have photometric and/or spectral variability characteristic of true LBVs \citep{Clark12,Lee14,Sholukhova14,Humphreys15}.  It is not clear what these cLBVs would have to do to be sanctified as true LBVs; the criteria needed seem to have been applied inconsistently and somewhat idiosyncratically within the literature.  For instance, \citet{Humphreys15} describes J004526.62+415006.3 as only the fifth ``confirmed" LBV in M31, despite the fact that other candidate LBVs have shown similar variability, such as J004051.59+403303.0 \citep{Sholukhova14}. 

\begin{figure}
\epsscale{0.48}
\plotone{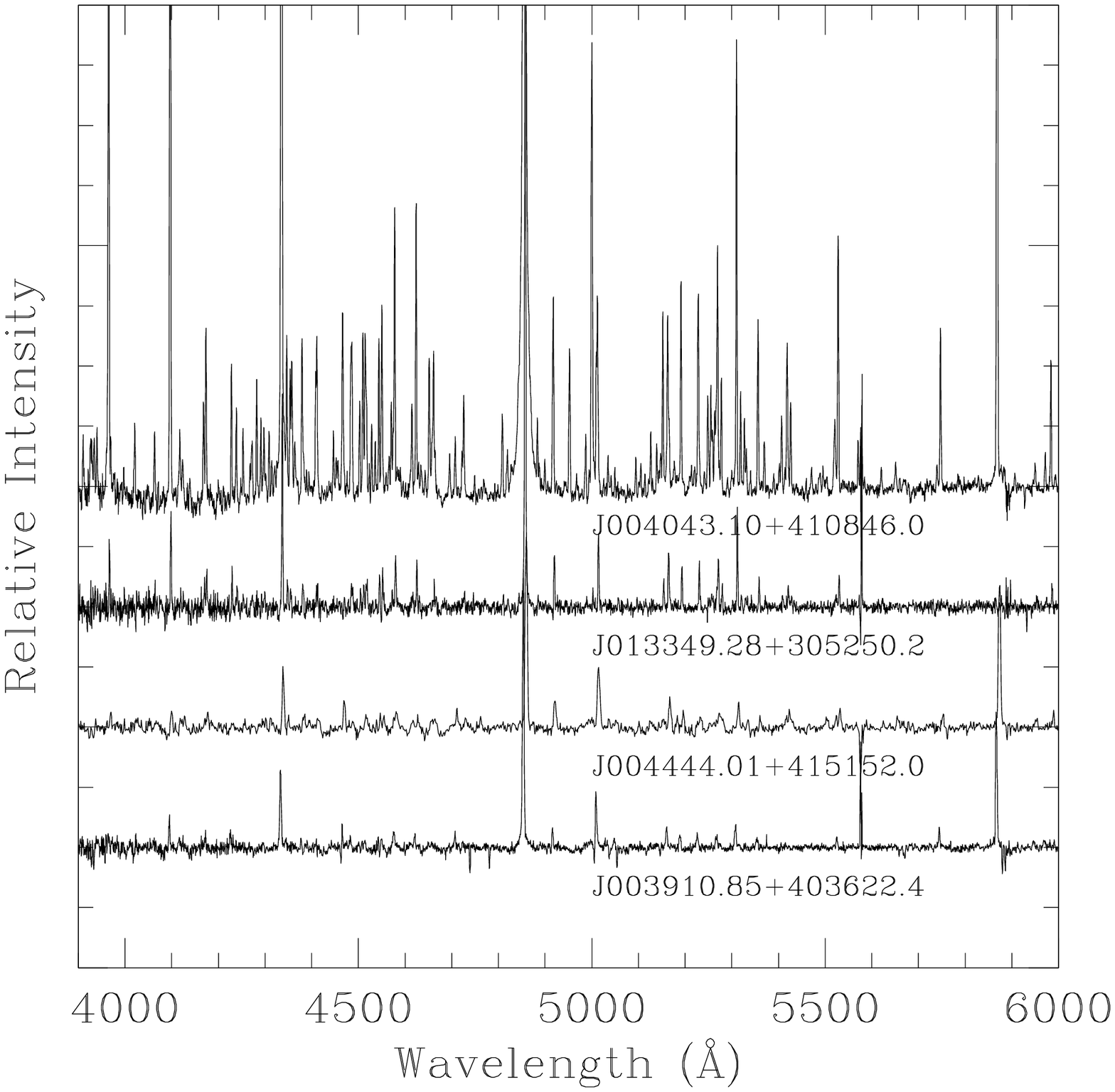}
\plotone{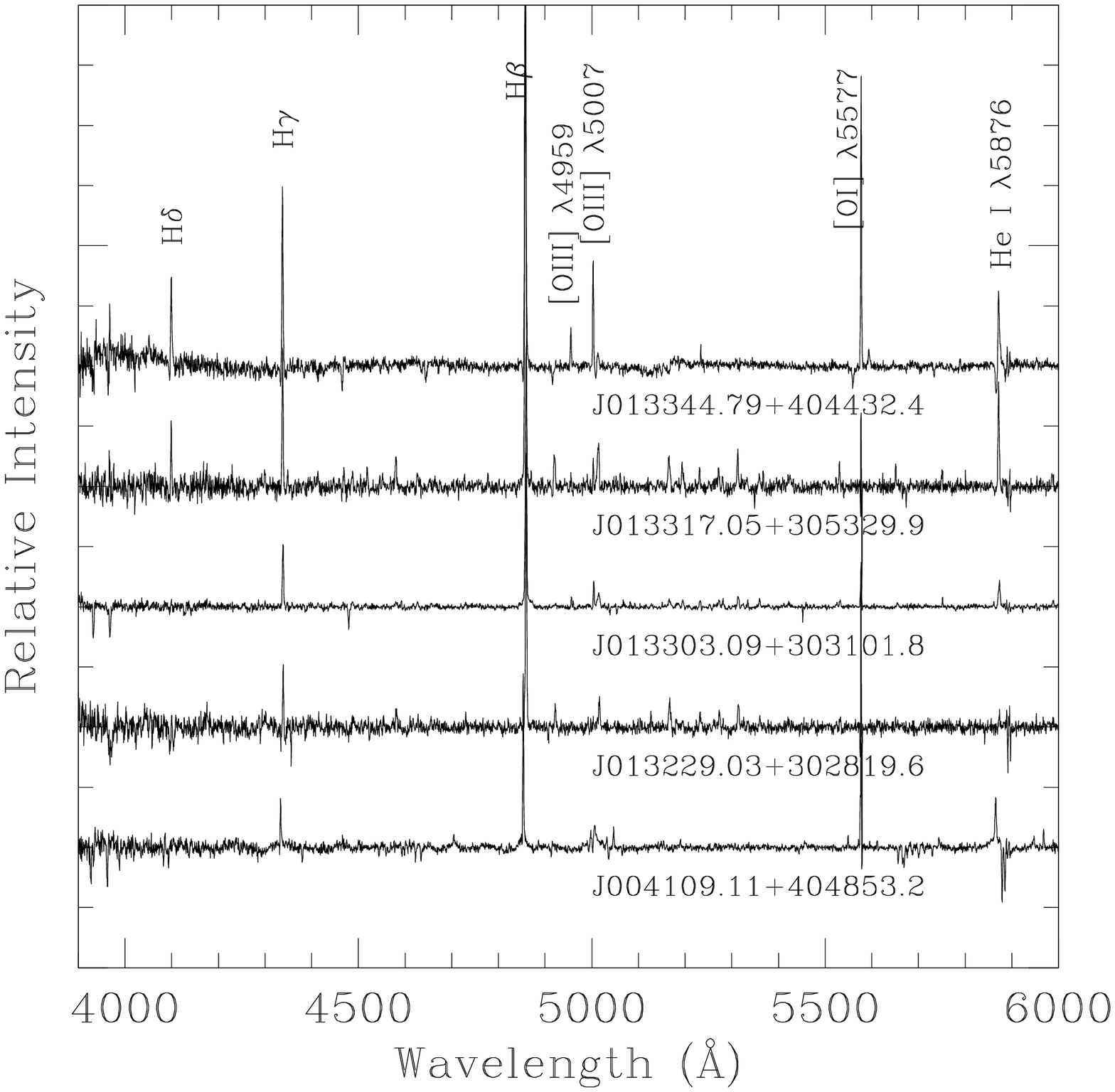}
\caption{\label{fig:LBVs} Newly discovered candidate LBVs (cLBVs) in M31 and M33.  On the left we illustrate the spectra of the strong-lined stars, and on the right we show the spectra of those with weaker emission. \cite[See, e.g.,][]{LGGSIII}.
The strong emission lines marked for the weak-lined stars could all be nebular in origin; what distinguishes these stars
from H~II regions (say) is the sea of Fe\,{\sc ii} and [Fe\,{\sc ii}] lines present.  These Fe lines dominate the spectrum for the strong-lined cLBVs.
}

\end{figure}

Our current spectroscopy has identified 9 newly found cLBVs in these two galaxies.    These new additions bring the number of known or suspected LBVs to 24 for M31 and 29 to M33\footnote{We note that \citet{Humphreys13} would call several of these ``post RSG warm hypergiants," unrelated to LBVs.  We have indicated these stars in Tables~\ref{tab:m31lbvs} and \ref{tab:m33lbvs}.  In addition, they consider J013232.86+303025.2 (one of the original  \citealt{HS} stars, Var~A), and J013415.42+302816.4 (which they refer to as ``N125093") as post RSG warm hypergiants.}.
In Figure~\ref{fig:LBVs} we illustrate the spectra of the newly found cLBVs, and we list all of the known or suspected LBVs in M31 and M33 in Tables~\ref{tab:m31lbvs} and \ref{tab:m33lbvs}.  With one exception, we exclude the so-called Ofpe/WN9 stars, now more commonly referred to as WN9-11 WRs \citep{CrowtherSmith97}. We realize that these late-type WNs stars have a strong linkage to LBVs.   In fact, \citet{NeugentM33} classified the star 
J013418.37+303837.0 as ``Ofpe/WN9" without realizing that this star is actually M33 Var 2, one of the early prototypes of the LBV class \citep{Humphreys78}. The lack of good coordinates for the confirmed LBVs is part of the reason for this oversight; a problem which we believe we have remedied here. To the best of our knowledge, the Var~2 has never before shown a late-type WN spectrum; \citet{Humphreys78} describes the star has having an A-F-type  spectrum along with Fe\,{\sc ii} and [Fe\,{\sc ii}].  Similarly, the star J013410.93+303437.6 was described as a newly found candidate LBV by \citet{LGGSIII}, when in fact this is the \citet{HS} star M33 Var 83.  

The star J0133395.2+304540.5 was classified as ``B0.5Ia+WNE" by \citet{NeugentM33}, but \citet{Clark12} has demonstrated strong spectral variability, and we remove it here from the list of WRs, and include it now as an LBV candidate. 

\begin{figure}
\epsscale{0.48}
\plotone{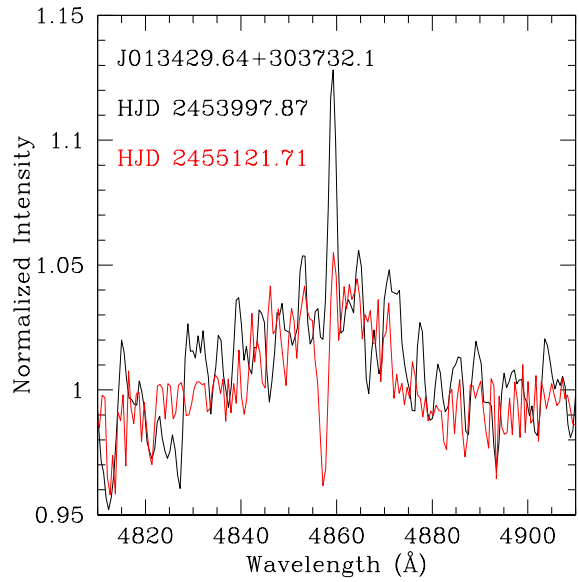}
\caption{\label{fig:B8I} The LBV candidate J013429.64+303732.1  The spectrum of the H$\beta$ profile has changed  in the spectrum of J013429.64+303732.1 over a 3 year interval.  Where there had been a narrow emission component superposed on a broad emission component, there is now narrow absorption; the broad component is unchanged.}.

\end{figure}

The star J013429.64+303732.1 was described by \citet{LGGSIII} as being a peculiar B8~I spectrum, showing narrow emission superimposed on a very broad component.  In Figure~\ref{fig:B8I} we show what appears to be changes that have taken place in the H$\beta$ profile over the three years after the \citet{LGGSIII} spectrum was taken: the broad emission component is unchanged, but there is now  an absorption component where there had been emission.  We cannot rule out an instrumental effect: the earlier spectrum was taken with a 3\arcsec diameter fiber with Hydra, while the later spectrum was taken with a 1\farcs5 fiber with Hectospec, each with different sky placements, so it is possible that nebular contamination is responsible for this change.  In addition, it is a minor change compared to the spectral variability found for true LBVs, and we leave its description as a ``candidate."

\begin{figure}
\epsscale{0.48}
\plotone{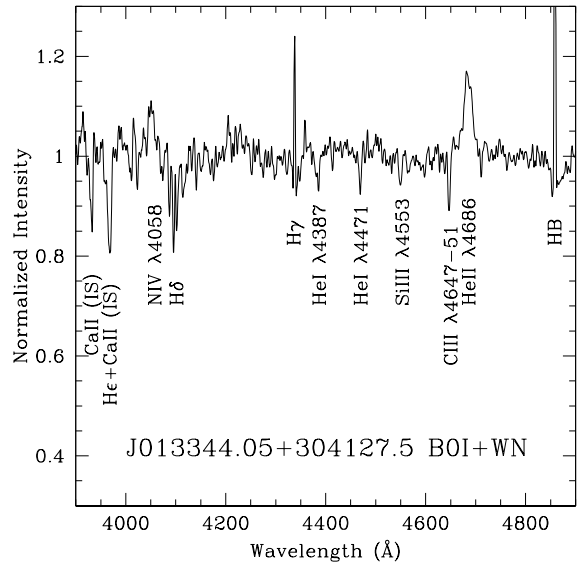}
\caption{\label{fig:newwr} The newly confirmed M33 WR star J013344.05+304127.5.  The spectrum of this star is dominated by the B0~I component, but with broad N~IV $\lambda 4058$ and He~II $\lambda 4686$ characteristic of a WN-type WR.}.

\end{figure}

\paragraph{\bf Wolf-Rayet Stars (WRs):}  \citet{NeugentM31} and \citet{NeugentM33} published the results of comprehensive surveys for WRs in M31 and M33, respectively.  Since that time \citet{SharaM31WR} found one
additional WR star in M31\footnote{We carefully examined the original WR filter images used by
\citet{NeugentM31}; the newly found WR is too heavily reddened and faint to have shown up in that survey.  In point of fact, the star is essentially invisible on the broad-band LGGS images as well.  It does not appear in the LGGS catalog.  \citet{SharaM31WR} use their discovery to argue that there could be a large
population of heavily reddened WRs that had been missed.  They based this argument in part on the fact that the number of WR stars in M31 is far fewer than the number known in the Milky Way.  Of course, for such scaling to be valid,  the current galaxy-wide star formation rates would have to be similar in the two galaxies. We believe it is
premature to extrapolate from one newly found reddened WR to a population of several thousand.}.
In addition, \citet{NeugentBinaries} spectroscopically confirmed five of the
previously unobserved M33 WR candidates listed by \citet{NeugentM33}.  (The sixth proved to be an Of-type star.) In the present paper we have identified one additional WR star in M33, J013344.05+304127.5, a B0~I+WN pair whose spectrum is dominated by the B0~I component.  We show the spectrum of this star in Fig.~\ref{fig:newwr}.  This brings the total number of WRs identified in M33 to 211, the 206 cataloged by \citet{NeugentM33}, minus J013395.2+304540.5 (which we now call a cLBV as discussed above), plus the addition of the five WRs confirmed by \citet{NeugentBinaries}, plus the one new one here.  All of the new additions are of WN-type. 

\paragraph{\bf O-type stars:}  All O-type stars in our selected magnitude range will be M31/M33 members, even the dwarfs.  The classification of O-type spectra are quite straightforward, relying upon the relative strengths of He\,{\sc i} and He\,{\sc ii}.  The presence of He\,{\sc ii} $\lambda 4686$ emission and/or N\,{\sc iii}~$\lambda \lambda 4634,42$ is then the primary luminosity indicator; more subtle aspects of this are described in both \citet{WF90} and \citet{Sota}. 

As Table~\ref{tab:newmembers} shows, most of the O-type stars known in M31 (83\%) and M33 (60\%) were identified as a result of our new spectroscopy here.  One of the stated goals of the current project was to improve our knowledge of the {\it unevolved} massive star populations of these galaxies, and while we have made an improvement, the number of O-type stars now classified in M31 (64) and in M33 (130) are only a few percent of the total number of expected O-type stars in these galaxies.  Current evolutionary models for solar metallicity predict there should be roughly 15$\times$ more O-type stars than WRs \cite[see Table 3 in][] {CyrilPops}.  Thus, we expect on the order 2000-3000 O type stars in each of these galaxies.  So, while we have made a substantial improvement, there is still a ways to go in terms of characterizing the unevolved massive star populations of these two galaxies\footnote{We were curious to see if this expected number of O-type actually existed in our photometry.  We therefore used the \citet{Sylvia} models to calculate the expected number of O stars in each magnitude interval, and compared that to the contents of the LGGS catalog.  The agreement was excellent, with the vast majority of the expected O stars to be found in the range $V=19.5-20.5.$  We will note that only about one-quarter of these O stars would be considered ``isolated" by our criteria.}.

\begin{figure}
\epsscale{0.48}
\plotone{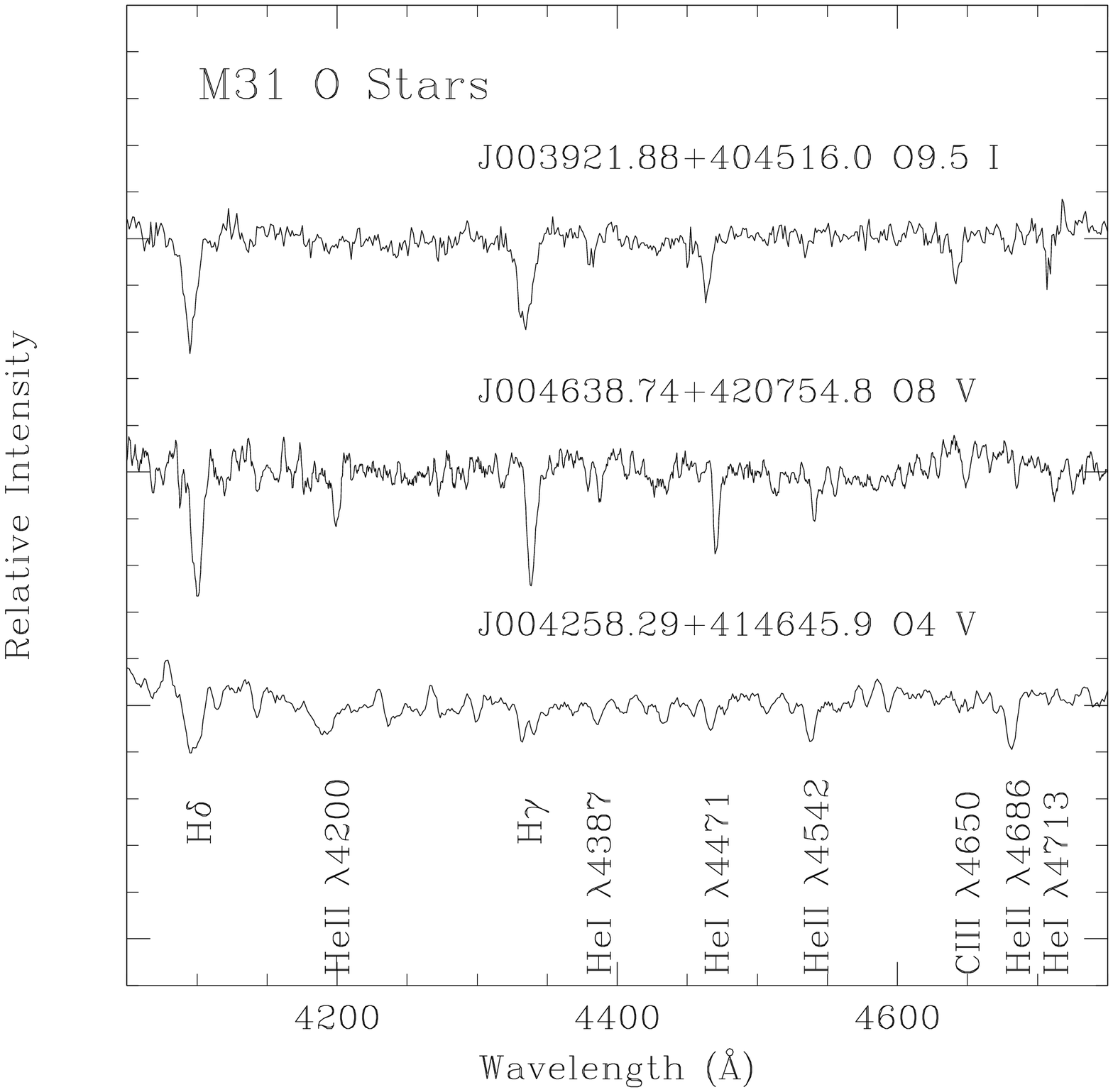}
\plotone{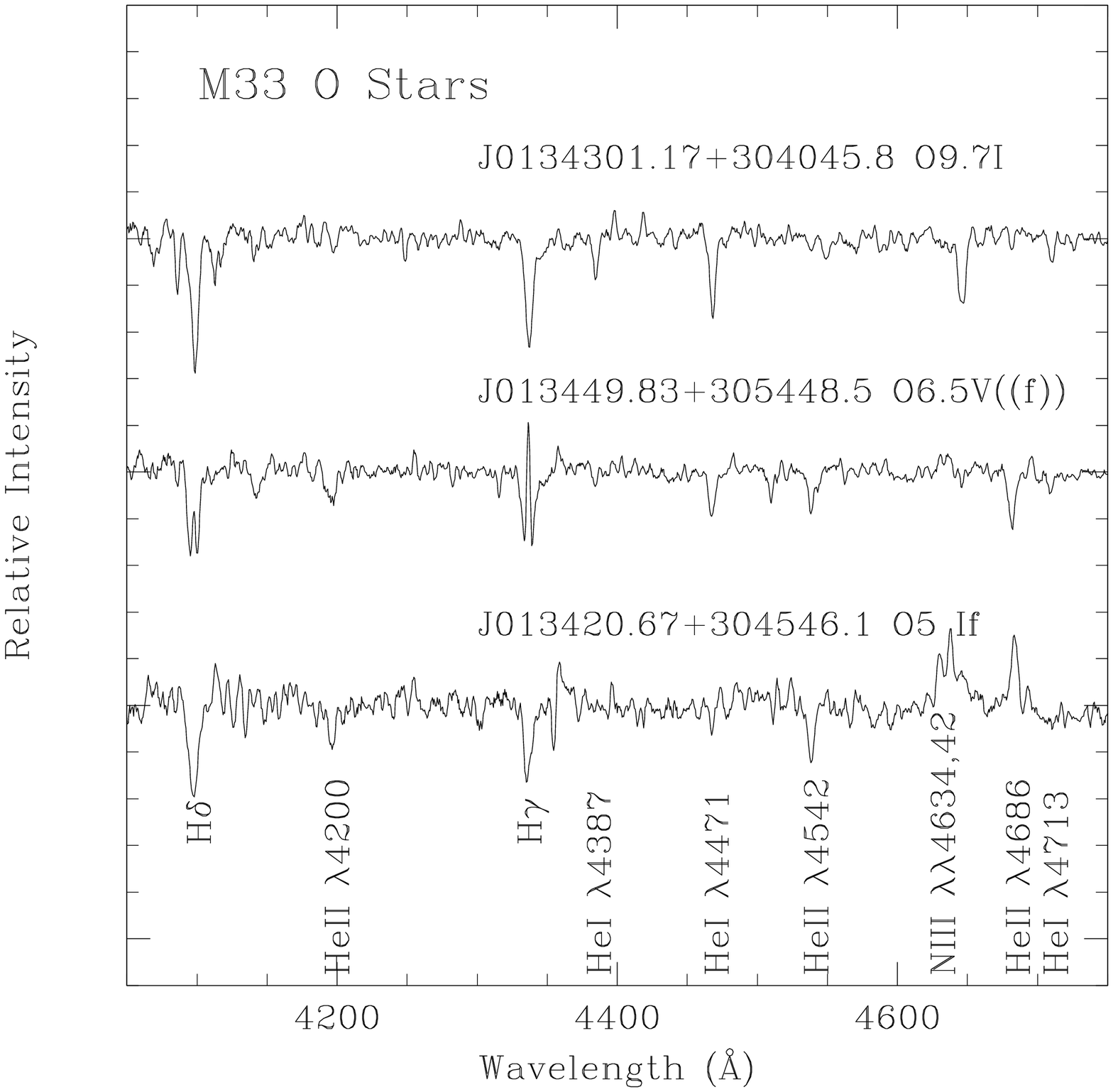}
\caption{\label{fig:Ostars} Examples of newly found O stars in M31 (left) and M33 (right).  J004258.29+414645.9 (O4~V) is the earliest type star known in M31, while J013413.06+305230.0 (O5~If) is the earliest known supergiant in M33. The classifications are based
on the relative strengths of He\,{\sc i} and He\,{\sc ii}.} 

\end{figure}

In Figure~\ref{fig:Ostars} we show examples of several of the newly found O-type stars in this study; the stars were picked as nice examples of early and late O-type dwarfs and supergiants.  We note that the O4~V star J004258.29+414645.9 is the earliest type star known in M31, while the O5~If star J013413.06+305230.0 is the earliest known supergiant in M33.  These spectra are remarkable mostly for their being so ordinary: they are typical of any similar stars in the Milky Way or Magellanic Clouds.  The only extraordinary fact is that these stars are $\sim$10$\times$ more distant than similar objects in the LMC or SMC.

\paragraph{\bf B-type Supergiants:} 
We expect only B-type supergiants and early B-type giants to be M31 members; at the same time, we expect no B-type foreground dwarfs to contaminate the sample.  The classification of B stars relies upon the relative strengths of Si\,{\sc iv}~$\lambda 4089$, Si\,{\sc iii}~$\lambda 4553$, and Si~II $\lambda 4128$, with the relative strengths of Mg\,{\sc ii}~$\lambda 4481$ and He\,{\sc i}~$\lambda 4471$ being a secondary indicator for B stars later than B2.  The luminosity criteria for the earliest B stars relies upon a comparison of Si\,{\sc iv} $\lambda 4116$ to He\,{\sc i} $\lambda 4121$, or Si\,{\sc iii} $\lambda 4553$ to He\,{\sc i} $\lambda 4387$.  At later B-types the strengths and shape of the Balmer lines were used to confirm a supergiant classification.

\begin{figure}
\epsscale{0.48}
\plotone{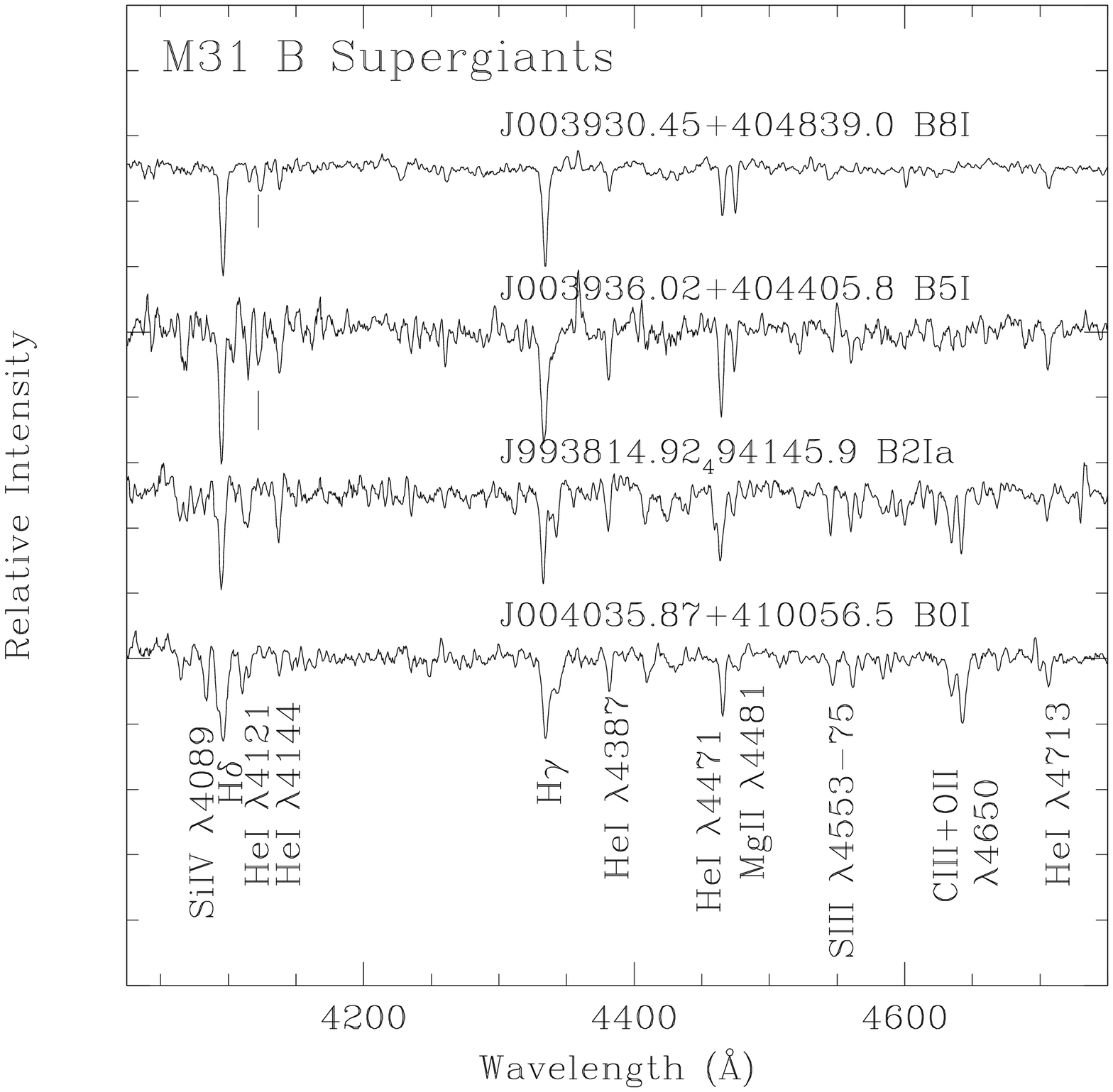}
\plotone{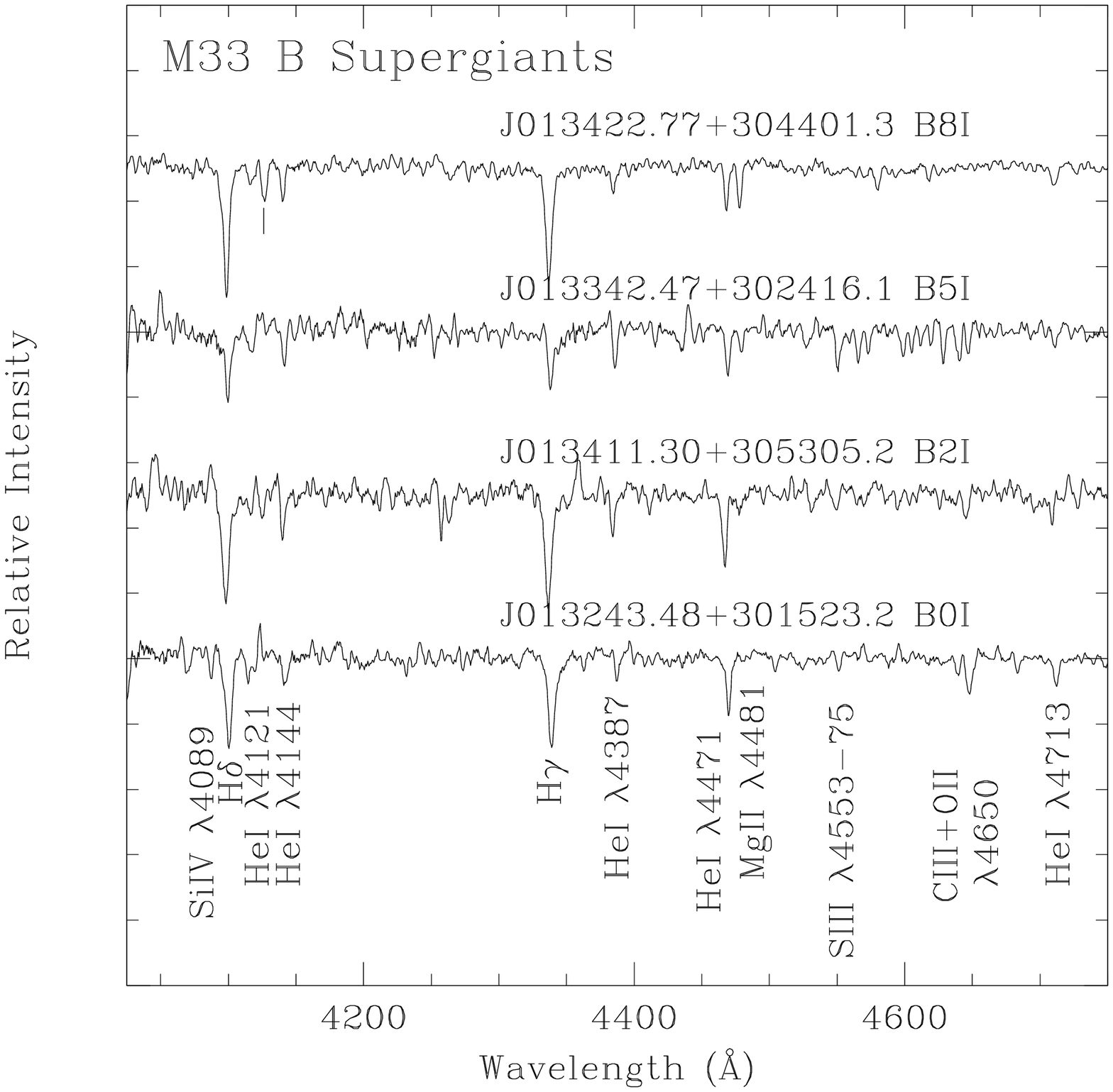}
\caption{\label{fig:Bstars} Examples of newly found B supergiants in M31 (left) and M33 (right). The Si\,{\sc ii} $\lambda 4128$ line is visible in the spectra of the later B-type supergiants between the two He I lines, as shown by the tick mark. The classifications are based primarily on the relative strengths of Si\,{\sc iv, iii,} and {\sc ii},  with the strength of Mg\,{\sc ii} being a secondary indicator.}

\end{figure}

Our study here is responsible for the vast majority (86\% for M31, and 79\%) of the B supergiants spectroscopically identified in M31 and M33, respectively (Table~\ref{tab:newmembers}).   We illustrate a few examples in Figure~\ref{fig:Bstars}\footnote{Our referee, Chris Evans, argues that the M33
star J013342.47+302416.1 should perhaps be classified a bit earlier than B5~I, as Si\,{\sc ii} is rather weak compared to a Si\,{\sc iii} for a B5~I type (compare to the B5~I M31 star illustrated next to it), although the He\,{\sc i} to Mg\,{\sc ii} ratio is consistent with a later type.  We believe this serves to illustrate how the spectral classification process is not an exact
one, especially in applying metal-dependent criteria (such as He\,{\sc i} vs. Mg\,{\sc ii}) in galaxies where the metallicity is substantially different than that where the spectral standards are defined.}.  

\paragraph{\bf A-type stars:} We expect some later-type A stars to be foreground dwarfs; fortunately the strengths of the Fe\,{\sc ii} and other metal lines, and the shape and strength of the Balmer lines readily distinguishes between the A-type supergiant members of M31/M33 and foreground dwarfs. The spectral subtypes are straightforwardly determined by comparing the strengths of the
Ca\,{\sc ii} H ($\lambda 3969$) and K ($\lambda 3934$) lines, as the former is blended with H$\epsilon$ and the strengths of the Balmer lines decrease as one goes from A0 to A9; an increase in the strengths of the metal lines also
was taken into account, as these reach a maximum at A5.

\begin{figure}
\epsscale{0.85}
\plotone{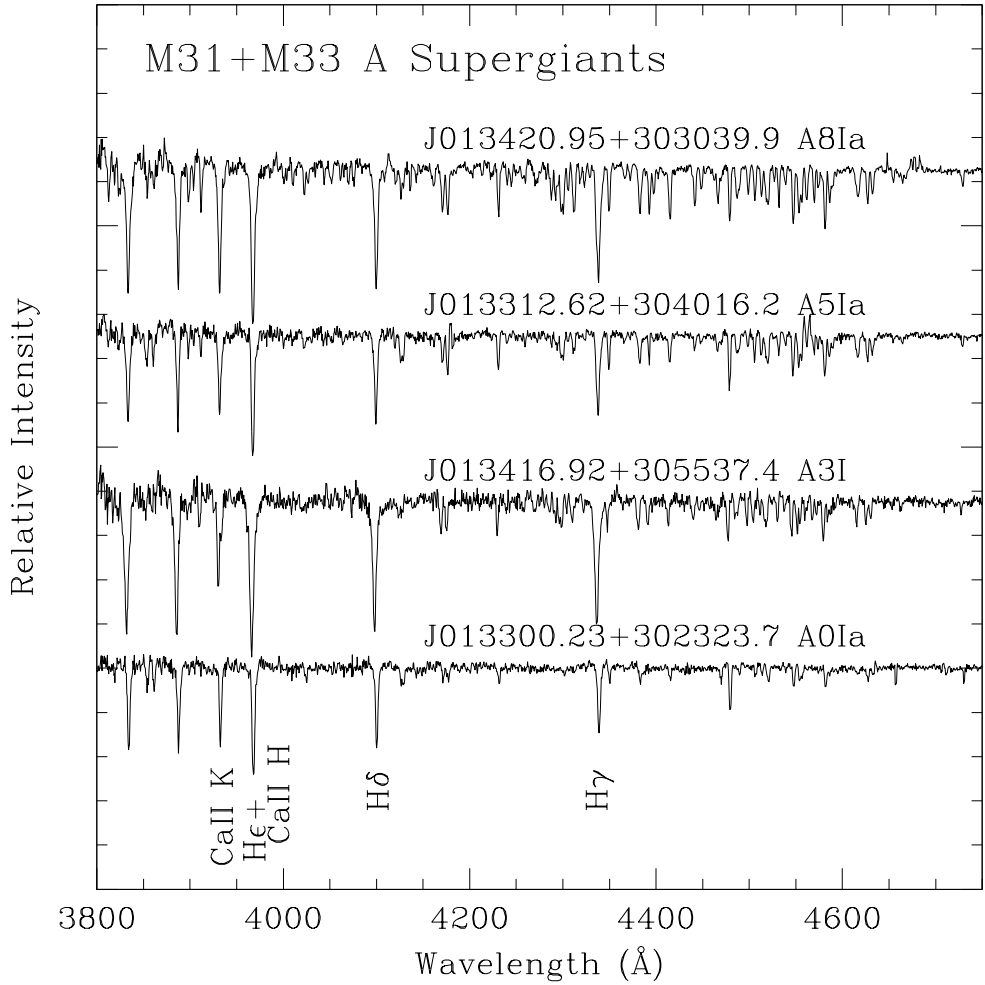}
\caption{\label{fig:Astars} Examples of newly found A supergiants in M31 and M33.  The shape and strengths of the Balmer lines make it easy to distinguish these supergiants from their foreground counterparts.}

\end{figure}

We see from Table~\ref{tab:newmembers} that about 90\% of the A-type supergiants known in M31 and M33 have been identified as a result of our new study.  We show some examples of A-type spectra in Figure~\ref{fig:Astars}. 

\paragraph{\bf F-type Supergiants:} Here we expect foreground dwarfs to dominate the sample (see, e.g., Figs.~\ref{fig:m31sel} and \ref{fig:m33sel}).  However, the F-type supergiants are very recognizable, as they have very strong metal lines. In addition, for F0-4, supergiants will have strong Ti\,{\sc ii}~$\lambda 4444$ relative to Mg\,{\sc ii}~$\lambda 4481$.  From F5-9, supergiants will not show the G-band, and Sr\,{\sc ii}~$\lambda 4077$ will have about half the strength of H$\delta$.  The spectral subtypes themselves were determined primarily by comparing the Ca\,{\sc i} $\lambda 4226$ line with H$\gamma$.

\begin{figure}
\epsscale{0.85}
\plotone{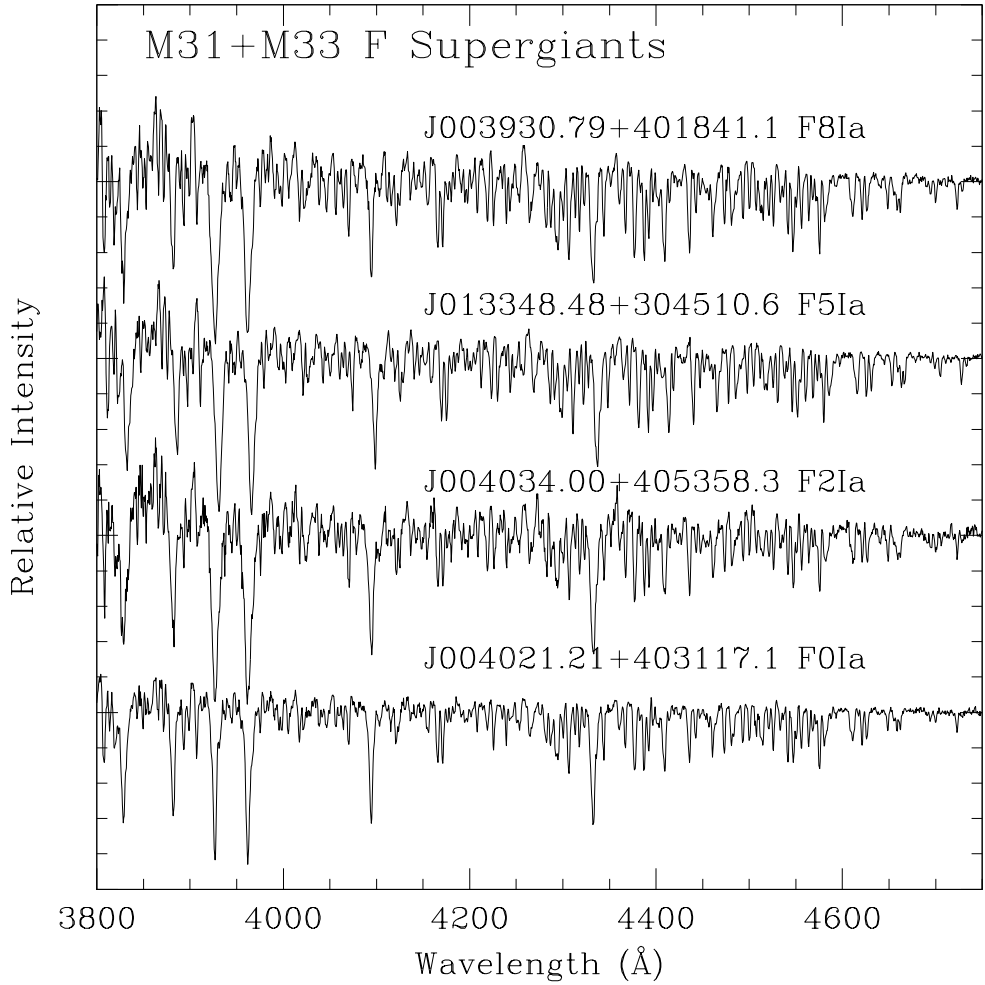}
\caption{\label{fig:Fstars} Examples of newly classified F supergiants in M31 and M33.  The  F-type supergiants are easily distinguished from foreground stars due to their rich metal-line spectra.}
\end{figure}

Most of the F-type supergiants known in these galaxies were discovered by the radial velocity surveys by \citet{DroutM31} and \citet{DroutM33}, and are classified here for the first time; most of the others were recently classified by \citet{Humphreys13}.  We show examples of our spectra in Figure~\ref{fig:Fstars}. 

\paragraph{\bf G-type Supergiants:} Again, foreground dwarfs will completely dominate the sample.  The occasional supergiant is easily recognized as having very sharp lines with the components of the G-band clearly separated.  In addition, supergiants have Sr\,{\sc ii}~$\lambda 4216$ comparable to that of Ca\,{\sc i}~$\lambda 4226$.  The spectral subtypes were determined by comparing H$\gamma$ and Fe\,{\sc i}~$\lambda 4325$, and H$\delta$ with Fe\,{\sc i}~$\lambda 4045$, with stronger Balmer lines indicating earlier types.

\begin{figure}
\epsscale{0.85}
\plotone{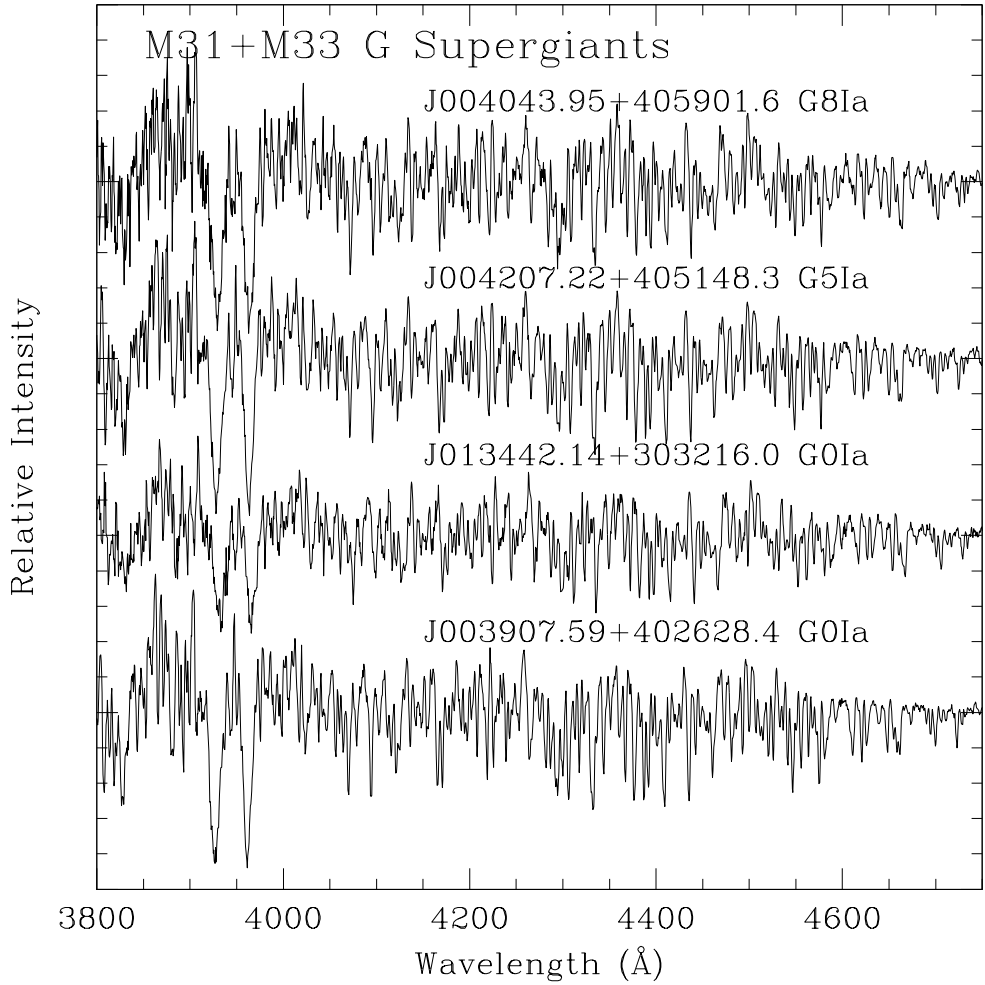}
\caption{\label{fig:Gstars} Examples of newly classified G supergiants in M31 and M33. The very sharp lines make these easily to distinguish from their foreground counterparts. }
\end{figure}

All of the known G-type supergiants in M31 and M33 have been classified as such as part of the present paper 
(Table~\ref{tab:newmembers}).  All but two of the M31 stars had been recognized as yellow supergiants by
the radial velocity surveys of \citet{DroutM31} and \citet{DroutM33}.   We show examples in Figure~\ref{fig:Gstars}.

\subsection{Location of Known Members in the CMD and HRD}

\begin{figure}
\epsscale{0.48}
\plotone{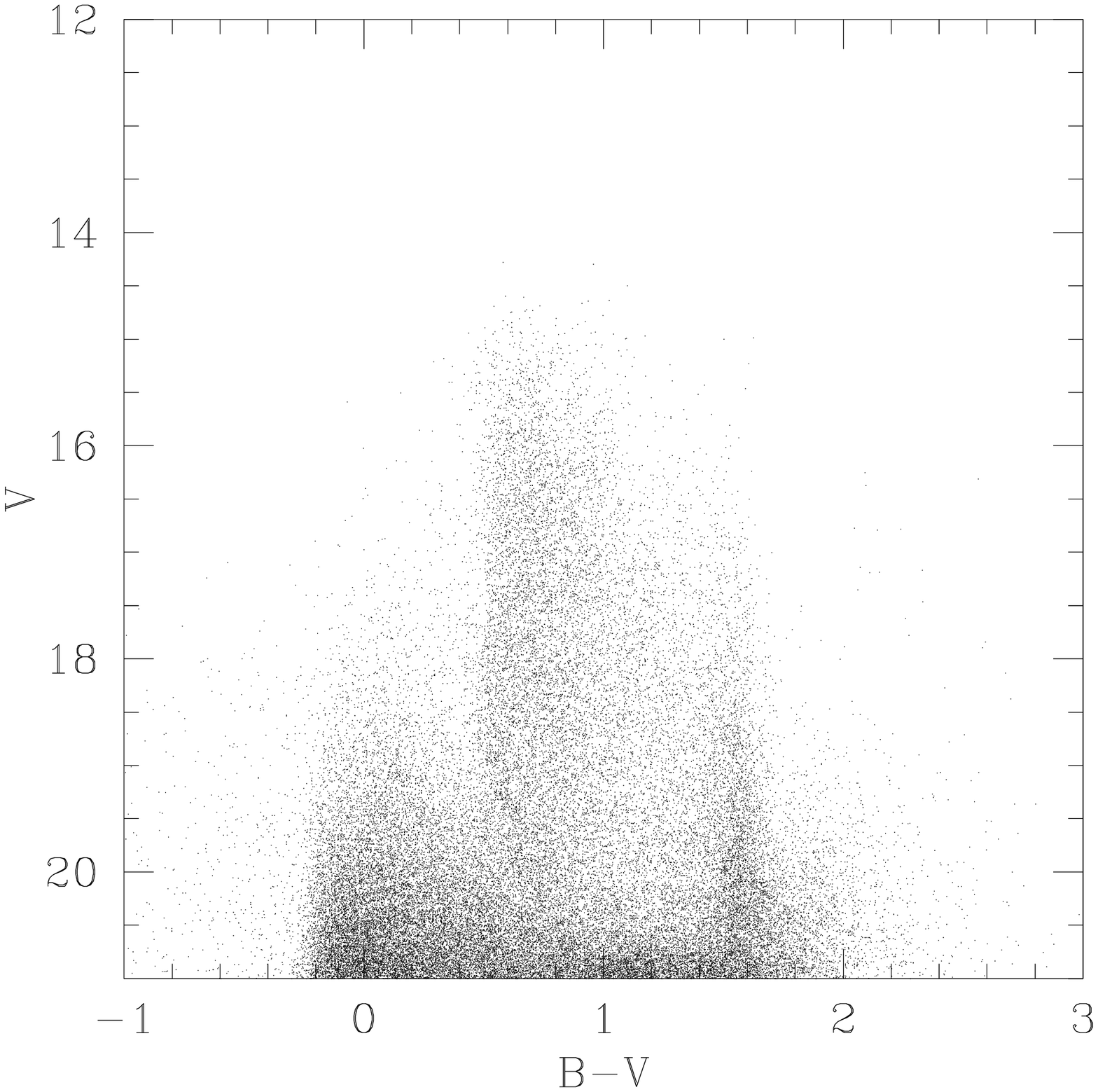}
\plotone{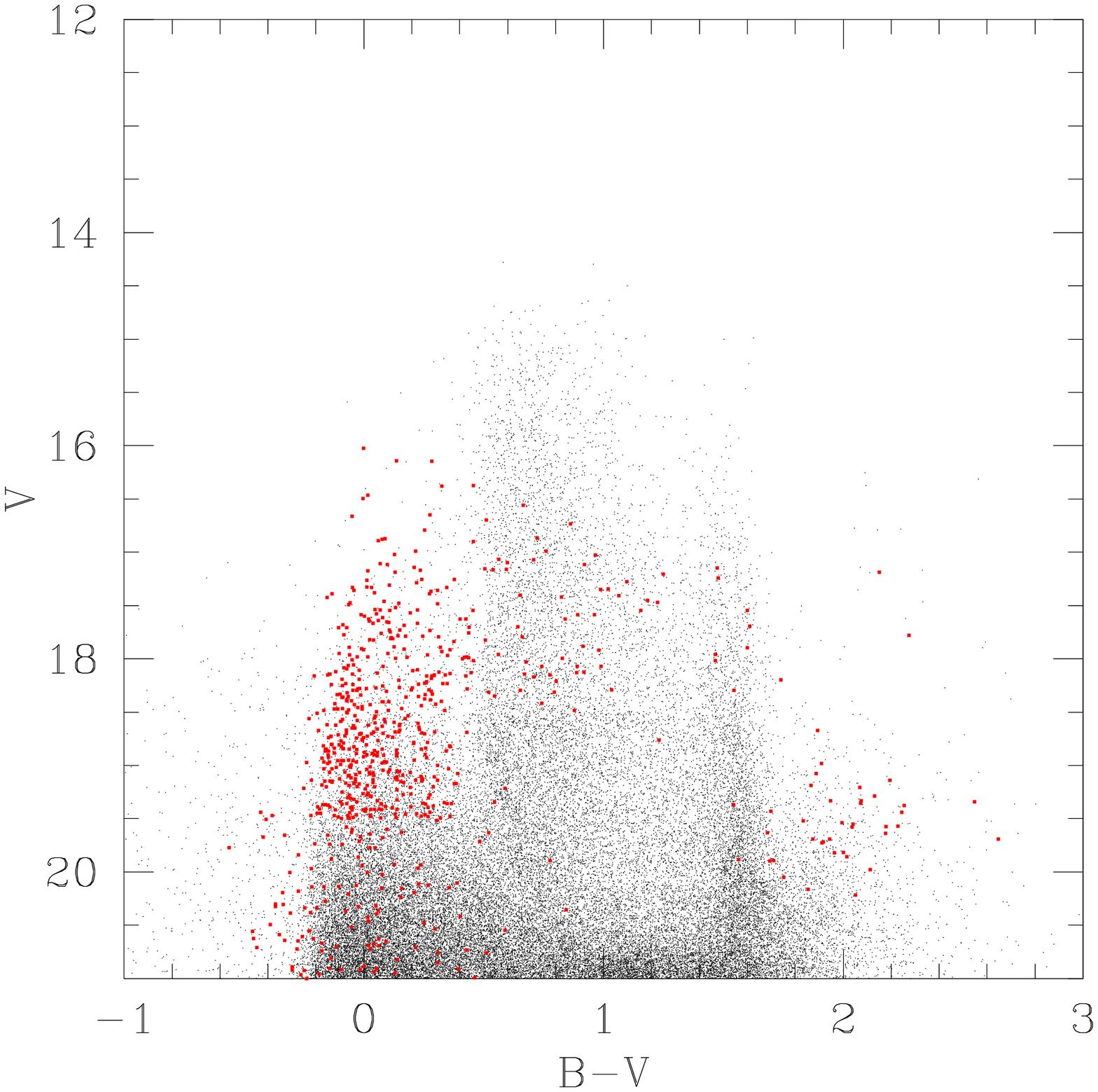}
\caption{\label{fig:M31cmdspect} Membership in the CMD for M31. {\it Left:}  We show a portion of the CMD for M31. {\it Right:} We now exclude stars known from our spectroscopy to be foreground stars, and indicate in red the stars known to be members.}
\end{figure}

\begin{figure}
\epsscale{0.48}
\plotone{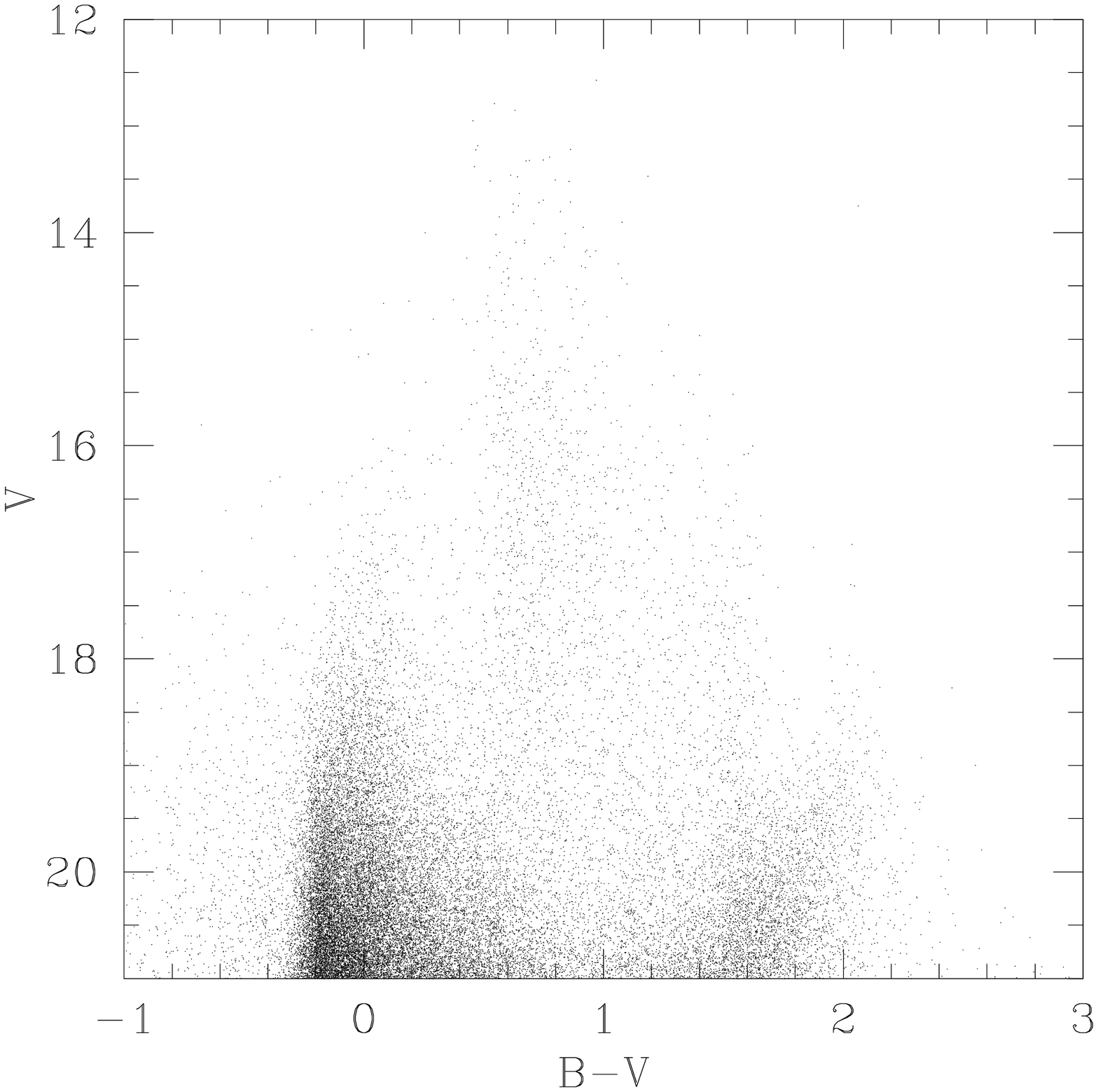}
\plotone{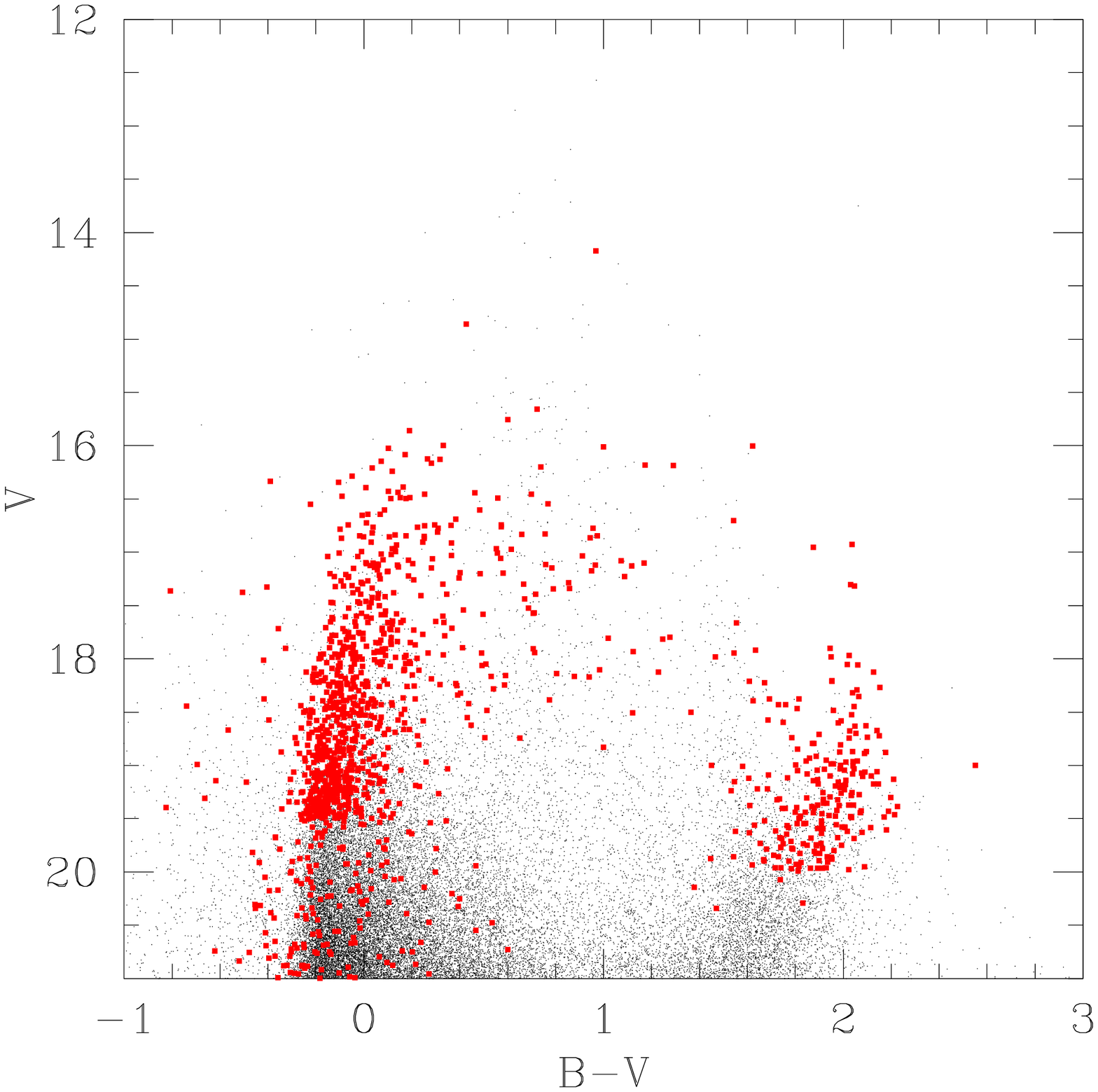}
\caption{\label{fig:M33cmdspect} Membership in the CMD for M33. {\it Left:}  We show a portion of the CMD for M33. {\it Right:} We now exclude stars known from our spectroscopy to be foreground stars, and indicate in red the stars known to be members.}
\end{figure}

In Figures~\ref{fig:M31cmdspect} and \ref{fig:M33cmdspect} we show the extent that our spectroscopy has affected
the interpretation of the color-magnitude diagrams.  On the left side we show the upper portion of the CMDs from 
Figures~\ref{fig:m31cmd} and \ref{fig:m33cmd}.  On the right, we have used the spectroscopic information now available to eliminate stars known to be foreground stars.  We see that the thick band of yellow stars has gotten thinner.  In addition, we indicate in red the stars for which our spectroscopy has established membership.  Primarily these
are stars in the ``blue plume," but with a smattering then of yellow and red supergiants. The contrast in how easily seen the red points are (denoting the stars with spectroscopy) demonstrates the fact that this work is considerably more advanced in M33 than in M31 right now (see also Table~\ref{tab:newmembers}),  a situation we hope to rectify in the coming years. 

Of course, we were most intrigued to see where these stars fell in a physical H-R diagram, with comparison to evolutionary
tracks.  Note, however, that there are strong selection effects: the numbers of WR stars are nearly complete \citep{NeugentM31,NeugentM33}, the numbers of yellow supergiants are probably complete 
\citep{DroutM31,DroutM33}, and
the number of RSGs is complete for M33 \citep{DroutM33} but not for M31 although work is in progress \citep{KateRunaway}.
At the same time, we have argued that the number of O-type stars known in these systems is woefully incomplete, and that our current knowledge represents only a few percent of the total.  {\it So, care must be exercised in not over interpreting these data.}
Still, it is of obvious interest to see where these fall in the HRD.

We have chosen to then plot only stars of spectral types O-F (and/or classified as YSGs), ignoring then the cLBVs, LBVs, the
WRs, and the RSGs.   The argument is that detailed modeling work is needed for even approximate temperatures for the (c)LBVs
and WRs, while the locations of RSGs require either model fitting or, at the least, K-band photometry, and all of this is best handled elsewhere.   We have also ignored stars that were obviously composite (double-lined or otherwise composite).  

For the O-G stars with MK types, we assigned effective temperatures based on their spectral types, using the effective temperature
scale of \citet{MasseyTeff} for the O stars and \citet{Allen} for B2~I through G8~I. (The B0-B2~I values were adjusted for a smooth transition.)    Photometry was corrected for reddening assuming $E(B-V)=0.13$ for M31 and  $E(B-V)=0.12$ for M33 \citep{LGGSII}; these values are averages, but the scatter for early-type stars is fairly small. The extinction at
$V$ was assumed then to be $A_V=3.1E(B-V)$.    
For the stars without MK types (such as those classified as YSGs) we used the de-reddened photometry and the relation  $$\log T_{\rm eff}=4.008-0.915(B-V)_0+0.9815(B-V)_0^2-0.36212(B-V)_0^3,$$  which is derived from a combination of model atmospheres \citep[see][]{DroutM33} and empirical data \cite[][and references therein]{Allen}, and is useful for supergiants over the range $3.6<\log T_{\rm eff} <4.2$ and $-0.1 <(B-V)_0 <1.6$\footnote{The relation given in \citet{DroutM33} is more precise, but restricted to the color and temperature range for yellow supergiants.}.  In M33 there were a few stars with ``generic" O-type designations that we also ignored.  The bolometric corrections were calculated using 
$$BC=-1131.425+827.397\times \log{T_{\rm eff}}-200.6348\times (\log{T_{\rm eff}})^2 + 16.11878\times\log{T_{\rm eff}}^3$$ based upon the values in \citet{MasseyTeff} and \citet{Allen}.  Distances of 760 and 830~kpc \citep{vandenbergh} were assumed for M31 and M33, respectively.

\begin{figure}
\epsscale{0.48}
\plotone{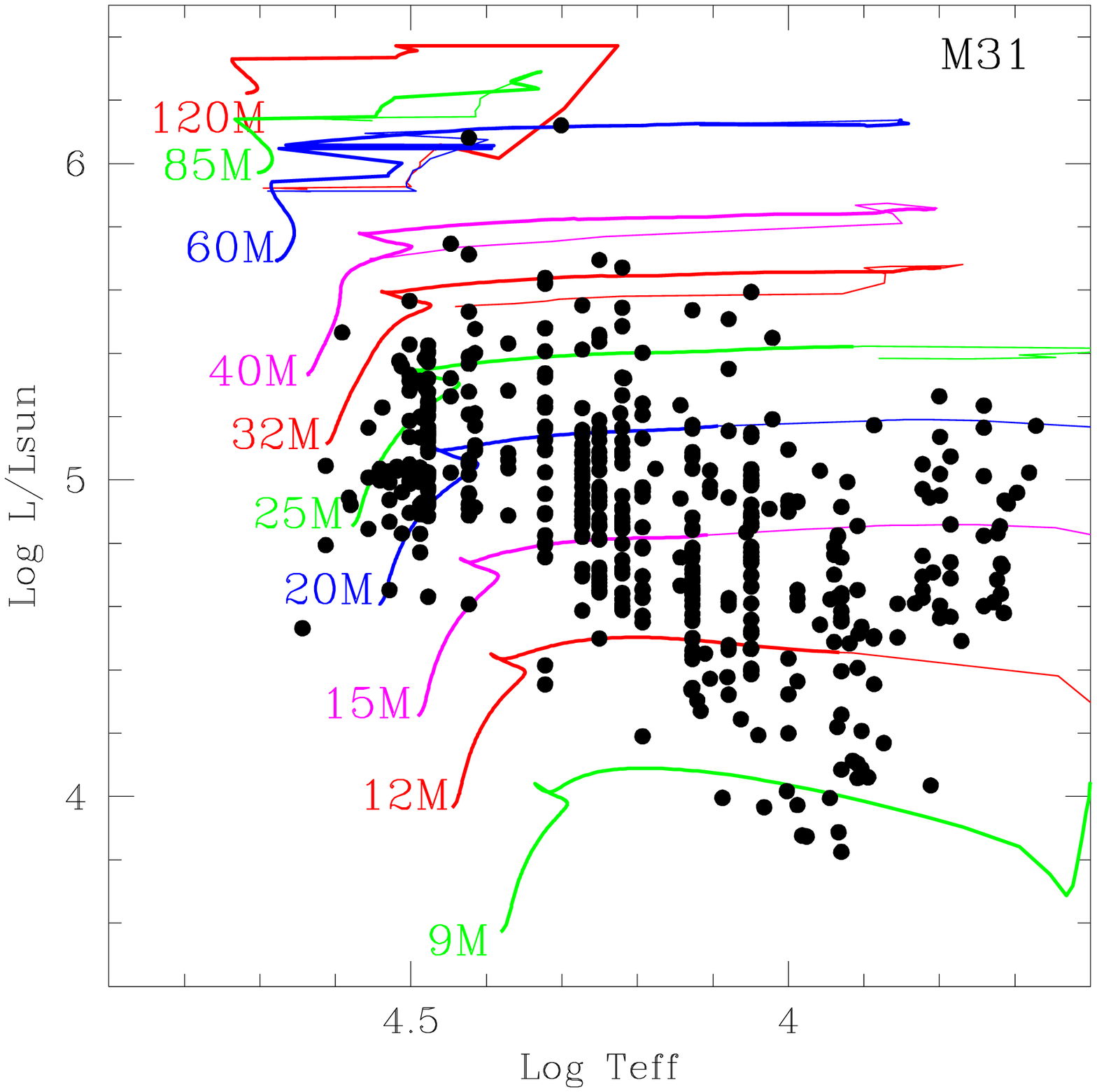}
\plotone{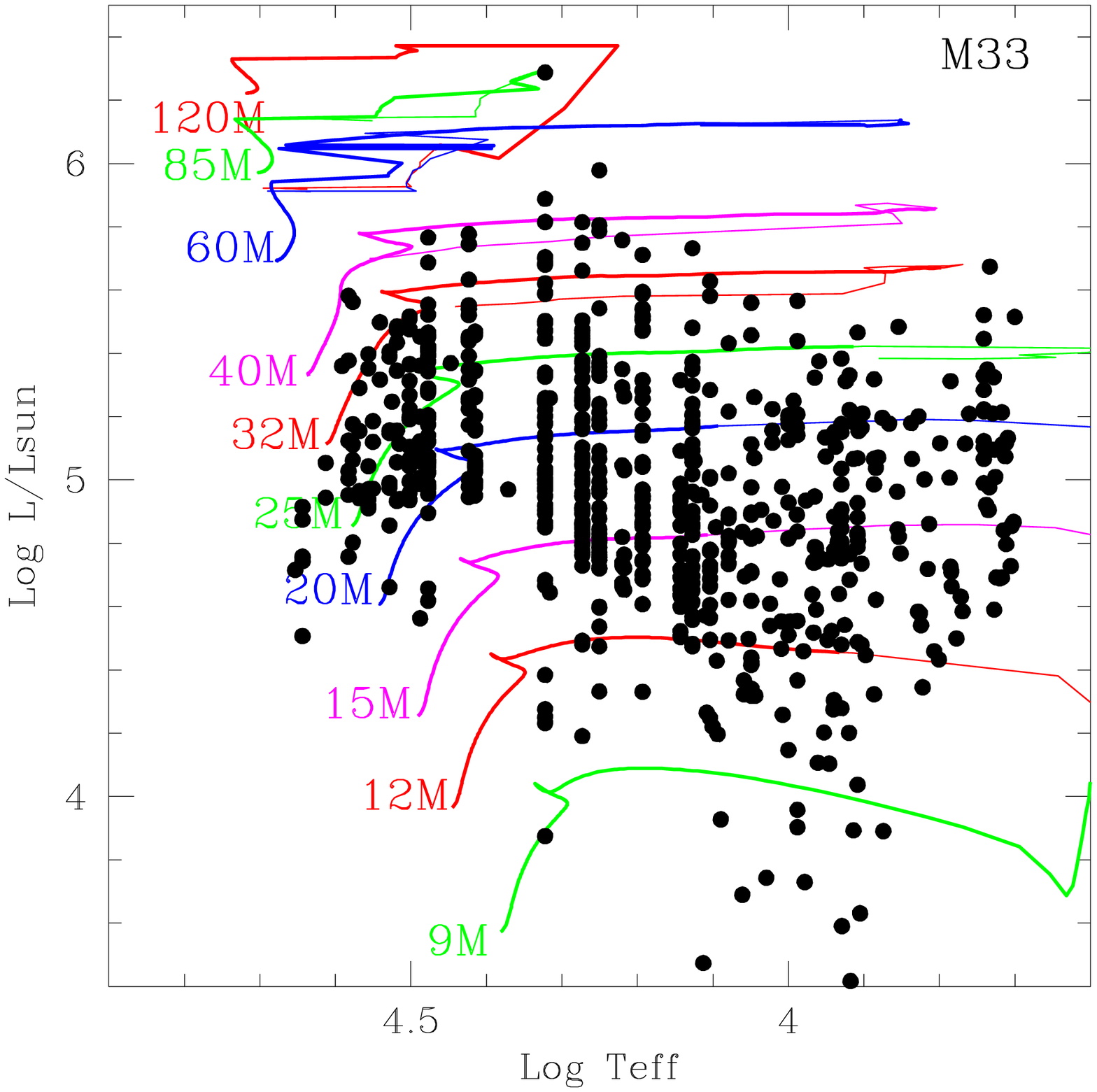}
\caption{\label{fig:HRDs} HRDs for known members in M31 (left)  and M33 (right).  The Geneva evolutionary tracks of \citet{Sylvia} are shown, with the initial masses indicated at the start of the tracks.  The bold part of the tracks indicate the H-burning phase. We have used the tracks with rotation (initial velocity 40\% of the breakup speed), and included only the first 209 points, corresponding to the beginning of any blue loop, for simplicity.}
\end{figure}

We show the resulting HRDs in Figure~\ref{fig:HRDs}.  We have included the solar metallicity evolutionary tracks from \citet{Sylvia}; these are computed for solar metallicity (Z=0.014) and thus are too low for M31 and too high for M33, but they give a reasonable idea of the mass ranges of the stars plotted.

The two highest mass stars in our sample for M31 are both early-type B supergiants, J004422.84+420433.1 (B0.5~I)
and J004057.86+410312.4 (B1-1.5~I).  These stars have masses of $\sim60M_\odot$ according to the evolutionary tracks.
While there are doubtless many O stars this massive or more so, they will be considerably fainter than these supergiants, owing to the larger (more negative) bolometric correction; it is for this reason that the younger massive stars are so disproportionately missing in our spectroscopic sample, a point that has
been emphasized many previous times \cite[see, e.g.,][]{MasseyLang,MasseyGilmore}: an ``old" (4~Myr) 40$M_\odot$ star will be a B0~I with $M_V=-6.6$, while the same star when it was younger (0-1~Myr) will be an O6~V that
is two magnitudes fainter visually! \citep[See Table~1 in][]{MasseyGilmore}.  The majority of the unevolved population we have identified in M31 are 20-40$M_\odot$.

The HRD of M33 is very similar.  Again, the highest luminosity (mass) star appears to are two early B supergiants, 
J013440.41+304601.4  (B1.5~I) and J013359.98+303354.7  (B1~Ie), and again the vast majority of the relatively unevolved massive stars in our sample are in the 20-40$M_\odot$ range.

\subsection{Revised LGGS Catalogs for M31 and M33}
In Tables~\ref{tab:M31BigTab} and \ref{tab:M33BigTab} we provide now the revised versions of the LGGS {\it UBVRI} catalogs of stars in M31 and M33.  The photometry is identical to that of \citet{LGGSI} with the {\it U-B} corrections as noted in the subsequent {\it erratum} \citep{LGGSIV}.  We have included not only the new spectral classifications derived here, but also updated the tables with all of the spectral types and information that have been published in the intervening years.  We have also included some other new information to make these tables more useful, including  membership criterion based upon the spectroscopy, an index indicating the degree of crowding, and a value for the galactocentric distance of each star within the plane of its galaxy.  

We have slightly modified the coordinates to place them essentially on the UCAC3/4 system \citep{UCAC4}; the original LGGS coordinates were tied to the USNO-B1.0 catalog \citep{USNOB1}, requiring small ($\sim$0\farcs3) corrections to place the stars on the same system as guide stars suitable for observing with Hectospec.     For M31:
$$\alpha_{\rm new}=\alpha_{\rm old} - 0.0065^{\rm s}$$ $$\delta_{\rm new}=\delta_{\rm old}+2\farcs283-3\farcs346\delta_{\rm old},$$ where the $\delta_{\rm old}$ in the last equation is expressed in radians.  For M33, \citet{NeugentM33} finds simpler corrections:
$$\alpha_{\rm new}=\alpha_{\rm old} - 0.030^{\rm s}$$ $$\delta_{\rm new}=\delta_{\rm old} - 0\farcs14.$$  The result of these corrections is that the LGGS designations no longer exactly correspond to the listed coordinates.

In addition, we have included an indication of how crowded based on the degree of contamination we expect in a 1\farcs5 fiber or slit under 1\farcs0 seeing conditions in either {\it B, V,} or {\it R}, whichever is worse.

\section{Summary and Future Work}

We have obtained new spectroscopy of 1895 stars in M31 and M33, 1496 of which are classified here for the first time.  This new spectroscopy is responsible now for the classification of 83\% (M31) and 60\% (M33) of the O stars,
87\% (M31) and 79\% (M33) of the B supergiants, 91\% (M31) and 87\% (M33) of the A-type supergiants, and 89\% (M31) and 76\% (M33) of the F-G supergiants now known in these two galaxies.  Our work has identified 9 new candidate LBV stars, and a new WR star.  

Nevertheless, this spectroscopy has admittedly only scratched the surface in characterizing the massive star populations of these systems.  We expect about $\sim 5000$ O-type stars to be present in total in these two galaxies  based upon the number of WRs known \citep{NeugentM31, NeugentM33}; this can be compared to the 194 O-type stars that have been classified.   Of course, in order to understand the massive star population, a complete spectroscopic survey may not be necessary, but it is also quite possible that there are subtypes 
of massive stars yet to be discovered.  We have recently seen an example of this in the Magellanic Clouds, where, despite decades of studies, a new class of WR stars has recently been found \citep{MasseyMCWRI,NeugentWN3O3} along with other unexpected objects \citep{MasseyMCWRII}.  The majority of the unevolved stars we have classified are in the mass range 20-32$M_\odot$ in both galaxies; the most luminous (and hence the
most massive) stars in our sample are early-type B supergiants.  This is to be expected, as more luminous O-type stars are visually fainter, and thus have not yet been observed. 

Of course, the classification of these stars is not an end in itself.  Rather, it will serve as a starting point for abundance studies of supergiants, investigations of the reddening laws in these galaxies, and provide the observational linchpins against which stellar evolutionary models can be evaluated.

\acknowledgements
Our good friend and colleague David R. Silva participated in some of the observing and made useful 
contributions at the beginning of the project. 
We are also indebted for the very fine support we received with Hectospec by
Perry Berlind, Mike Calkins, and Marc Lacasse, the excellent MMT operators Erin Martin, John McAfee, and 
Alejandra Milone, and much useful advice and good queue scheduling by Nelson Caldwell.  Deidre Hunter was kind enough to make useful comments on an earlier draft of the manuscript.  Our referee, Chris Evans, made a number of valuable suggestions which have helped us improve this paper.
This project was partially supported by the National Science Foundation by grant AST-1008020, as well as by Lowell Observatory.  

{\it Facility:} \facility{MMT (Hectospec)}

\clearpage

\bibliography{lgsurveyads}

\clearpage

\clearpage

\clearpage

\clearpage

\clearpage

\begin{deluxetable}{l c c}
\tablecaption{\label{tab:sample}Photometric Criteria Used}
\tablewidth{0pt}
\tablehead{
\colhead{Parameter}
&\multicolumn{2}{c}{Criterion} \\ \cline{2-3} 
&\colhead{$V\leq 18.0$} & \colhead{$19.5\geq V > 18.0$}
}
\startdata
$B-V$    & $\leq 0.4$  & $\leq 0.4$\\
$U-B$  &   $\leq 0.05-1.125(B-V)$ & $\leq -0.4$ \\
$Q$       &   $\leq 0.0$   & $\leq -0.7$  \\
\enddata
\end{deluxetable}

\begin{deluxetable}{l r r r r r}
\tablecaption{\label{tab:newmembers}Confirmed M31/M33 Members}
\tablewidth{0pt}
\tablehead{
\colhead{Type}
&\multicolumn{2}{c}{M31}
&&\multicolumn{2}{c}{M33} \\ \cline{2-3} \cline{5-6}
&\colhead{New}
& \colhead{Total}
&&\colhead{New}
& \colhead{Total}
}
\startdata
(c)LBVs & 4 & 25 && 5 & 32 \\
WRs & 0 & 154 && 1 & 211 \\
O~V, III, I & 53 & 64 && 78 & 130  \\
B~III, I & 278 & 321 && 371 & 471  \\
A~I & 39 & 43 && 62 & 71 \\
F~I & 21 & 25 && 6 & 11   \\
G~I & 10 & 10 && 10 & 10 \\
K~I & 0 & 7 && 0 & 2 \\
M~I  & 1 & 38 && 2 & 7  \\
YSGs & 0 & 14 && 0 & 86 \\
RSGs & 0  & 15 && 0 & 220 \\
Star+Neb & 3 & 14 && 18 & 57 \\
\enddata
\end{deluxetable}

\clearpage
\begin{deluxetable}{l r r r r r c c l l }
\tabletypesize{\scriptsize}
\setlength{\tabcolsep}{3pt}
\tablecaption{\label{tab:m31lbvs}M31 (c)LBVs}
\tablewidth{0pt}
\tablehead{
\colhead{Star}
&\colhead{$V$}
&\colhead{$B-V$}
&\colhead{$U-B$}
&\colhead{$V-R$}
&\colhead{$R-I$\tablenotemark{a}}
&\colhead{Cwd\tablenotemark{b}}
&\colhead{Ref\tablenotemark{c}}
&\colhead{Type}
&\colhead{Cross-ID}
}
\startdata
J003910.85+403622.4  & 18.181  & 0.282  & -1.045   & 0.368   & 0.269  & I & 1    & cLBV \\
J004043.10+410846.0  & 18.616  & 0.150  & -1.063   &1.058  & -0.162  & I & 1    & cLBV \\
J004051.59+403303.0  & 16.989  & 0.216  & -0.761   & 0.220   & 0.193  & X& 2    & cLBV \\
J004109.11+404853.2  & 18.151  & 0.294  & -0.833   & 0.274   & 0.259  & I & 1    & cLBV: \\
J004221.78+410013.4  & 19.632  & 0.522  & -0.340   &1.184   & 0.034  & I & 3    & cLBV \\
J004229.87+410551.8  & 18.785  & 0.266  & -0.628   & 0.558  & -0.206  & C& 4   & cLBV &\\
J004302.52+414912.4  & 17.426  & 0.153  & -0.895   & 0.184   & 0.001  & I & 7   & LBV &  AE And\\
J004320.97+414039.6  & 19.215  & 0.589  & -0.826   & 0.768   & 0.474  & I & 6   & cLBV &  k114a\\
J004322.50+413940.9  &  20.354 & 0.844 & -0.503  & 0.511  & 0.737 & I & 4 & cLBV\tablenotemark{d} \\
J004333.09+411210.4  & 17.325  & 0.013  & -0.975   & 0.153  & -0.085  & I & 5   & LBV &  AF And\\
J004341.84+411112.0  & 17.547  & 0.455  & -0.760   & 0.407   & 0.287  & I & 8   & cLBV &\\
J004350.50+414611.4  & 17.700  & 0.642  & -0.356   & 0.471   & 0.482  & I & 4   & cLBV &\\
J004411.36+413257.2  & 18.071  & 0.742  & -0.493   & 0.616   & 0.496  & I & 6   & cLBV &  k315a\\
J004415.00+420156.2  & 18.291  & 0.272  & -0.793   &1.061  & -0.089  & I & 4   & cLBV &\\
J004417.10+411928.0  & 17.113  & 0.099  & -0.718   & 0.337   & 0.166  & I & 6   & cLBV &  k350\\
J004419.43+412247.0  & 18.450  & 0.007  & -0.569   & 0.539   & 0.052  & I & 5   & LBV &  Var 15\\
J004425.18+413452.2  & 17.479  & 0.147  & -0.916   & 0.216   & 0.107  & I & 6   & cLBV &  k411\\
J004442.28+415823.1  & 19.679  & 0.078  & -1.189   & 0.621  & -0.018  & I & 4   & cLBV &\\
J004444.01+415152.0  & 19.034  & 0.013  & -1.101   & 0.174   & 0.123  & I & 1    & cLBV &\\
J004444.52+412804.0  &  18.073 & 0.989 & -0.084 & 0.747 & 0.765 & I & 4 & cLBV\tablenotemark{d} \\
J004450.54+413037.7  & 17.143  & 0.211  & -0.658   & 0.353   & 0.149  & C& 5   & LBV &  Var A-1\\
J004507.65+413740.8  & 16.145  & 0.283  & -0.520   & 0.211  & 99.999  & X& 4   & cLBV &\\
J004522.58+415034.8  & 18.467  & 0.158  & -0.720   & 0.272   & 0.175  & I & 4   & cLBV\tablenotemark{d} &\\
J004526.62+415006.3  & 17.157  & 0.505  & -0.908   & 0.318   & 0.129  & X& 4   & cLBV &\\
J004621.08+421308.2  & 18.155  & 0.294  & -0.573   & 0.434   & 0.332  & I & 6   & cLBV &  k895\\
\enddata
\tablenotetext{a}{99.999=no value.}
\tablenotetext{b}{Crowding: I=isolated (contamination $<$5\%), S=somewhat isolated (5-20\%), C=crowded (20-80\%), X=extremely crowded ($>$80\%).}
\tablenotetext{c}{References for spectral type: 
1=Current paper;
2=\citealt{LGGSI};
3=\citealt{NeugentM31};
4=\citealt{LGGSIII};
5=\citealt{HS,Humphreys75};
6=\citealt{King};
7=\citealt{ST,Humphreys75};
8=\citealt{M31PCyg}.
}
\tablenotetext{d}{Considered a ``post-RSG warm hypergiant" by \citealt{Humphreys15}.}
\end{deluxetable}

\clearpage
\begin{deluxetable}{l r r r r r c c l l }
\tabletypesize{\scriptsize}
\setlength{\tabcolsep}{3pt}
\tablecaption{\label{tab:m33lbvs}M33 (c)LBVs}
\tablewidth{0pt}
\tablehead{
\colhead{Star}
&\colhead{$V$}
&\colhead{$B-V$}
&\colhead{$U-B$}
&\colhead{$V-R$}
&\colhead{$R-I$\tablenotemark{a}}
&\colhead{Cwd\tablenotemark{b}}
&\colhead{Ref\tablenotemark{c}}
&\colhead{Type}
&\colhead{Cross-ID}
}
\startdata
J013229.03+302819.6 & 18.998 &  0.041 & -0.774  & 0.127  & 0.108& I&  1 &   cLBV:\\
J013235.25+303017.6 & 18.007 &  0.045 & -0.959  & 0.136  & 0.165 &I & 2  &  cLBV\\
J013242.26+302114.1 & 17.440  & 0.671 & -1.070  & 0.890 & 99.999 &C  &2 &   cLBV\\
J013248.26+303950.4 & 17.252 &  0.068 & -1.191  & 0.240  & 0.127& I & 2   & cLBV\\
J013300.02+303332.4  &18.317 & -0.125 & -1.088 & -0.055 & -0.017& I & 3  &  cLBV    & UIT026 \\
J013303.09+303101.8 & 16.994 &  0.132 & -0.734  & 0.141  & 0.122& I & 1   & cLBV\\
J013317.05+305329.9 & 18.921  & 0.042 & -1.100  & 0.255  & 0.076 &I&  1  &  cLBV\\
J013324.62+302328.4 & 19.579 & -0.085 & -0.838  & 0.309 & -0.256& I  &2  &  cLBV\\
J013332.64+304127.2 & 18.985 & -0.060 & -0.842 &  0.155 &  0.110& I & 2  &  WNL/cLBV   \\        
J013333.22+303343.4  &19.397 & -0.101 & -0.991  & 0.958 & -0.449 &I & 2   & cLBV\\
J013335.14+303600.4&  16.429 &  0.102 & -0.988 &  0.132 &  0.161& X & 4 &  LBV     & Var C  \\  
J013339.52+304540.5 & 17.503 &  0.064 & -1.014  & 0.099  & 0.054 &I & 5 &  cLBV    \\
J013341.28+302237.2 & 16.285  &-0.049 & -1.055  & 0.006 & -0.042 &I & 2  & cLBV    & 101-A \\
J013344.79+304432.4&  18.151 &  0.202  &-0.763  & 0.221  & 0.182 &I  &1   & cLBV:\\
J013349.23+303809.1 & 16.208 &  0.035  &-0.913  & 0.182 &  0.108& X&  4   &LBV     & Var B \\   
J013349.28+305250.2 & 19.043 &  0.155  &-1.042  & 0.321 &  0.089 &I & 1    &cLBV\\
J013350.12+304126.6 & 16.819 &  0.035&  -1.095  & 0.385  & 0.128 &X & 3   & cLBV    & UIT212 \\
J013351.46+304057.0  &17.728 &  0.077 & -0.927  & 0.128  & 0.089& C & 2,6 & cLBV: \\
J013352.42+303909.6 & 16.165 &  0.281 & -0.441  & 0.204  & 0.249 &X&  7 &  cLBV\tablenotemark{d}\\
J013355.96+304530.6 & 14.859&   0.428 & -0.346  & 0.314 &  0.147 &I  &5 &  cLBV\tablenotemark{d}    & UIT247, B324 \\
J013357.73+301714.2 & 17.387  & 0.060  &-0.649 &  0.097  & 0.086& I & 2 & cLBV\\
J013406.63+304147.8 & 16.084 &  0.173  &-1.139  & 0.226 &  0.097 &I&  3    &cLBV    & UIT301 \\
J013410.93+303437.6&  16.027  & 0.103  &-1.004 &  0.154  & 0.172& I & 8  & LBV     & Var 83 \\
J013416.10+303344.9 & 17.120  & 0.050  &-0.852  & 0.156 &  0.117& X & 3,9& cLBV    & UIT341 \\
J013416.44+303120.8  &17.096 &  0.009& -1.029  & 0.058 &  0.036& I & 2   & cLBV\\
J013418.37+303837.0 & 18.219 & -0.129 & -1.025 &  0.120  & 0.089& X  & 4,6 &LBV/Ofpe/WN9 & Var 2 \\
J013422.91+304411.0 & 17.215 &  0.065 & -0.925  & 0.070&   0.071 &X & 2    &cLBV\\
J013424.78+303306.6 & 16.837&   0.131 & -0.753 &  0.115 & 0.179& I  &2  & cLBV \\
J013426.11+303424.7 & 18.969 &  0.259  &-0.845 &  0.383  & 0.321 &I  &2   & cLBV\\
J013429.64+303732.1 & 17.105 &  0.010  &-0.871 &  0.055&   0.119 &I & 1,2  &B8Ipec/cLBV & \\
J013459.47+303701.9 & 18.374 &  0.213 & -1.234  & 0.490  & 0.194 &I & 2 &   cLBV\\
J013500.30+304150.9 & 19.298 & -0.073&  -0.854 &  0.698 & -0.509 &I & 2 &   cLBV\\
\enddata
\tablenotetext{a}{99.999=no value.}
\tablenotetext{b}{Crowding: I=isolated (contamination $<$5\%), S=somewhat isolated (5-20\%), C=crowded (20-80\%), X=extremely crowded ($>$80\%).}
\tablenotetext{c}{References for spectral type: 
1=Current paper;
2=\citealt{LGGSIII};
3=\citealt{UIT};
3=\citealt{NeugentM31};
4=\citealt{HS,Humphreys75};
5=\citealt{Clark12};
6=\citealt{NeugentM33};
7=\citealt{Valeev};
8=\citealt{Humphreys78};
9=\citealt{Monteverde96}.
}
\tablenotetext{d}{Considered a ``post-RSG warm hypergiant" by \citealt{Humphreys15}.}
\end{deluxetable}

\clearpage

\thispagestyle{empty}

\begin{deluxetable}{l rrrrrrrrrrrrrccccccccl}
\tabletypesize{\tiny}
\setlength{\tabcolsep}{3pt}
\rotate
\tablecaption{\label{tab:M31BigTab}Updated M31 LGGS Catalog}
\tablewidth{0pt}
\tablehead{
\colhead{Star}
&\colhead{$\rho$\tablenotemark{a}}
&\colhead{$\alpha_{\rm 2000}$}
&\colhead{$\delta_{\rm 2000}$}
&\colhead{$V$}
&\colhead{$\sigma_V$}
&\colhead{$B-V$}
&\colhead{$\sigma_{\rm B-V}$}
&\colhead{$U-B$\tablenotemark{b}}
&\colhead{$\sigma_{\rm U-B}$\tablenotemark{b}}
&\colhead{$V-R$}
&\colhead{$\sigma_{\rm V-R}$}
&\colhead{$R-I$\tablenotemark{b}}
&\colhead{$\sigma_{\rm R-I}$\tablenotemark{b}}
&\colhead{$N_V$}
&\colhead{$N_B$}
&\colhead{$N_U$}
&\colhead{$N_R$}
&\colhead{$N_I$}
&\colhead{Cwd\tablenotemark{c}}
&\colhead{Mem\tablenotemark{d}}
&\colhead{Ref\tablenotemark{e}}
&\colhead{Type}
}
\startdata              
J003701.92+401233.2 & 1.24 &00 37 01.91& +40 12 33.1 &19.862  &0.017& -0.021 & 0.021 &-0.928 & 0.015 & 0.204&  0.023 &99.999 &99.999  &1  &2  &1 & 1  &0& I      \\                 
J003701.93+401218.4  &1.24 &00 37 01.92& +40 12 18.3 &18.739&  0.008 & 1.494 & 0.015&  0.945 & 0.036 & 0.946 & 0.014 &99.999 &99.999 & 1 & 2 & 1 & 1 & 0 &I       \\                
J003702.03+401141.4 & 1.23 &00 37 02.02 &+40 11 41.3 &21.225 & 0.043 & 1.362 & 0.085 &99.999 &99.999 & 0.748 & 0.049  &0.694  &0.024 & 1 & 1 & 0  &1 & 1& I   \\                    
J003702.05+400633.5  &1.18 &00 37 02.04 &+40 06 33.4 &21.091&  0.044 & 0.050 & 0.061& -1.110&  0.052 & 0.042 & 0.074 &99.999 &99.999  &1 & 2  &1  &1 & 0& I    \\                   
J003702.13+400945.6 & 1.21 &00 37 02.12 &+40 09 45.5& 16.091 & 0.006 & 1.287&  0.007&  0.983 & 0.007&  0.792 & 0.010 &99.999& 99.999  &1  &2  &1 &1  &0 & I &0 &1   & fgd    \\        
J003702.24+401225.7 & 1.24 &00 37 02.23 &+40 12 25.6 &20.765 &0.029 & 1.584&  0.072 &99.999 &99.999&  1.030 & 0.036  &1.262  &0.021  &1  &2  &0  &1 & 1& I      \\                 
J003702.38+400529.5 & 1.18 &00 37 02.37 &+40 05 29.4& 22.427  &0.165 & 1.359 & 0.498 &99.999 &99.999 & 0.574&  0.180  &0.892 & 0.072 & 1 & 1 & 0  &1 & 1& I   \\                    
J003702.44+400723.2 & 1.19 &00 37 02.43 &+40 07 23.1 &22.461  &0.173& -0.153&  0.202 &99.999& 99.999& -0.018 & 0.252 &99.999 &99.999 & 1 & 2 & 0 & 1  &0 &I  \\                    
J003702.47+401742.5 & 1.30 &00 37 02.46& +40 17 42.4 &18.026 & 0.007 & 1.324&  0.011&  1.150 & 0.023 & 0.732 & 0.011 & 0.688  &0.008  &1 & 2 & 1  &1  &1 &I &0 &1 &   fgd     \\       
J003702.51+401654.5&  1.29 &00 37 02.50 &+40 16 54.4 &18.768 & 0.008  &0.687&  0.012& -0.040 & 0.014 & 0.393 & 0.011 & 0.390 & 0.008  &1 & 2 & 1&  1 & 1 &I  \\    
\enddata  
\tablecomments{Table~\ref{tab:M31BigTab} is published in its entirety in the electronic edition.
A portion is shown here for guidance regarding its form and content.}
\tablenotetext{a}{Deprojected distance within the plane of M31 in units of the Holmberg radius.  Assumes 
a Holmberg radius of 95\farcm3, inclination 77\fdg0, and position angle of major axis of 35\fdg0. At a distance of 760 kpc, a value of 1.00 for $\rho$ corresponds to 21.07~kpc.}
\tablenotetext{b}{99.999=no value.}
\tablenotetext{c}{Crowding: I=isolated (contamination $<$5\%), S=somewhat isolated (5-20\%), C=crowded (20-80\%), X=extremely crowded ($>$80\%).}
\tablenotetext{d}{Membership: 0=non-member, 1=member, 2=possible member, 3=unknown}
\tablenotetext{e}{References for spectral type: 
1=\citealt{DroutM31};
2=Current paper;
3=\citealt{LGGSI};
4=\citealt{Trundle};
5=\citealt{Humphreys88};
6=\citealt{Humphreys79};
7=\citealt{NeugentM31};
8=\citealt{MasseySilva};
9=\citealt{CordinerIII};
10=\citealt{MasseyRSG};
11=\citealt{HumpMassey};
12=\citealt{CordinerI};
13=\citealt{MasseyWilson};
14=\citealt{LGGSIII};
15=\citealt{Humphreys13};
16=\citealt{MJ98};
17=\citealt{HS};
18=\citealt{King};
19=\citealt{M31PCyg};
20=\citealt{MasseyEvans}.
}
\end{deluxetable} 

\clearpage
\thispagestyle{empty}

\begin{deluxetable}{l rrrrrrrrrrrrrccccccccl}
\tabletypesize{\tiny}
\setlength{\tabcolsep}{3pt}
\rotate
\tablecaption{\label{tab:M33BigTab}Updated M33 LGGS Catalog}
\tablewidth{0pt}
\tablehead{
\colhead{Star}
&\colhead{$\rho$\tablenotemark{a}}
&\colhead{$\alpha_{\rm 2000}$}
&\colhead{$\delta_{\rm 2000}$}
&\colhead{$V$}
&\colhead{$\sigma_V$}
&\colhead{$B-V$}
&\colhead{$\sigma_{\rm B-V}$}
&\colhead{$U-B$\tablenotemark{b}}
&\colhead{$\sigma_{\rm U-B}$\tablenotemark{b}}
&\colhead{$V-R$}
&\colhead{$\sigma_{\rm V-R}$}
&\colhead{$R-I$\tablenotemark{b}}
&\colhead{$\sigma_{\rm R-I}$\tablenotemark{b}}
&\colhead{$N_V$}
&\colhead{$N_B$}
&\colhead{$N_U$}
&\colhead{$N_R$}
&\colhead{$N_I$}
&\colhead{Cwd\tablenotemark{c}}
&\colhead{Mem\tablenotemark{d}}
&\colhead{Ref\tablenotemark{e}}
&\colhead{Type}
}
\startdata              
J013146.18+302931.4 & 1.37 &01 31 46.15& +30 29 31.3 &21.027 & 0.061&  0.090 & 0.113&  0.266 & 0.118 & 1.012 & 0.111& 99.999 &99.999 & 1 & 1 & 1 & 1&  0 &X\\
J013146.18+302932.4  &1.37& 01 31 46.15& +30 29 32.3 &20.560  &0.068 & 0.645&  0.117 &99.999 &99.999 & 0.564 & 0.115& 99.999 &99.999 & 1 & 1&  0 & 1 & 0 &C\\
J013146.20+302706.2 & 1.36 &01 31 46.17& +30 27 06.1 &21.057&  0.032 & 1.857 & 0.084& 99.999 &99.999  &0.924 & 0.036 &99.999& 99.999&  1&  1 & 0 & 1&  0& I\\
J013146.21+302026.9 & 1.36 &01 31 46.18& +30 20 26.8& 21.179 & 0.038  &0.962 & 0.066  &0.749&  0.096&  0.588 & 0.047& 99.999 &99.999 & 1 & 1 & 1 & 1 & 0& I\\
J013146.25+301849.7 & 1.36& 01 31 46.22& +30 18 49.6 &16.359 & 0.005 & 0.811 & 0.007 & 0.470&  0.007 & 0.460&  0.007 &99.999 &99.999&  1  &1&  1 & 1  &0& I& 0 &1&    fgd\\
J013146.26+302931.5  &1.37 &01 31 46.23& +30 29 31.4 &20.247  &0.063  &0.912 & 0.128 & 0.223  &0.131& -0.209 & 0.112 &99.999 &99.999 & 1  &1  &1  &1 & 0 &C\\
J013146.43+302048.4&  1.35 &01 31 46.40 &+30 20 48.3& 21.990  &0.074&  0.460 & 0.102& -0.155 & 0.089  &0.308&  0.097 &99.999& 99.999 & 1&  1 & 1&  1 & 0& I\\
J013146.64+301756.1 & 1.36& 01 31 46.61& +30 17 56.0 &22.075  &0.080 & 1.498 & 0.156 &99.999& 99.999  &1.121 & 0.086 &99.999 &99.999  &1  &1 & 0  &1  &0 &I\\
J013146.73+303118.0 & 1.37 &01 31 46.70 &+30 31 17.9 &22.110 & 0.080 & 1.256&  0.153& 99.999 &99.999 & 0.710 & 0.088 & 0.805  &0.037  &1 & 1 & 0 & 1  &1& I\\
J013146.81+302614.2 & 1.35& 01 31 46.78& +30 26 14.1 &22.258  &0.094 & 1.856 & 0.274& 99.999 &99.999  &0.952&  0.104&  1.115 & 0.044 & 1 & 1 & 0&  1&  1& I\\

\enddata  
\tablecomments{Table~\ref{tab:M31BigTab} is published in its entirety in the electronic edition.
A portion is shown here for guidance regarding its form and content.}
\tablenotetext{a}{Deprojected distance within the plane of M33 in units of the Holmberg radius.  Assumes 
a Holmberg radius of 30\farcm8, inclination 56\fdg0, and position angle of major axis of 23\fdg0. At a distance of 830 kpc, a value of 1.00 for $\rho$ corresponds to 7.44~kpc.}
\tablenotetext{b}{99.999=n o value.}
\tablenotetext{c}{Crowding: I=isolated (contamination $<$5\%), S=somewhat isolated (5-20\%), C=crowded (20-80\%), X=extremely crowded ($>$80\%).}
\tablenotetext{d}{Membership: 0=non-member, 1=member, 2=possible member, 3=unknown}
\tablenotetext{e}{References for spectral type: 
1=\citealt{DroutM33};
2=Current paper;
3=\citealt{U};
4=\citealt{NeugentM33};
5=\citealt{Humphreys13};
6=\citealt{Humphreys80};
7=\citealt{MJ98};
8=\citealt{LGGSIII};
9=\citealt{UIT};
10=\citealt{MasseyRSG};
11=\citealt{MasseyWilson};
12=\citealt{LGGSI};
13=\citealt{NeugentBinaries};
14=\citealt{Drissen08};
15=\citealt{vandenBergh75};
16=\citealt{Abbott04};
17=\citealt{Monteverde96};
18=\citealt{Kehrig};
19=\citealt{HS};
20=\citealt{CordinerII};
21=\citealt{OeyMassey};
22=\citealt{Valeev};
23=\citealt{Clark12};
24=\citealt{Humphreys78};
25=\citealt{Urb}.}
\end{deluxetable}

\end{document}